\documentclass[12pt]{article}
\usepackage{latexsym,graphicx,amssymb,color,amsmath}
\usepackage{epsfig}
\usepackage{lscape}
\date{}
\graphicspath{}
\DeclareGraphicsExtensions{.eps}
\newtheorem{lemma}{\bfseries Lemma}[section]

\newtheorem{theorem}[lemma]{\bfseries Theorem}

\newcommand\fr[2]{{\textstyle\frac{#1}{#2}}}

\newcommand{\be}{\begin{equation}}
\newcommand{\ee}{\end{equation}}
\newcommand{\barray}[1][rcl]{\be\begin{array}{#1}}
\newcommand{\barrayl}[2][rcl]{\be\label{#2} \begin{array}{#1}}
\newcommand{\earray}{\end{array}\ee}

\def\K{K_{{\rm d}_{app}}}

\newenvironment{proof}{\medskip\par\noindent{\scshape Proof:}}%
 {\hfill{$\Box$}\bigskip\par}%

\begin{document}

\vspace*{-2cm}
\begin{flushright}
DTP--05/09\\
\end{flushright}

\vspace{0.3cm}

\begin{center}
{\Large {\bf Assembly Models for Papovaviridae based on \\ \vspace{0.3cm}
Tiling Theory }}\\ 
\vspace{1cm} {\large \bf T.\ Keef\,\footnote{\noindent E-mail: 
{\tt tk506@york.ac.uk}}, A.\
Taormina\,\footnote{\noindent E-mail: {\tt anne.taormina@durham.ac.uk}} and
R.\ Twarock\,${}^{1,}$\footnote{\noindent E-mail: 
{\tt rt507@york.ac.uk}}}\\
\vspace{0.3cm} {${}^1$}\em Department of Mathematics \\ University of York\\
\vspace{0.3cm} {${}^3$} Department of Biology \\ University of York \\
York YO10 5DD, U.K.\\ \vspace{0.3cm} {${}^2$\em \it Department of Mathematical
Sciences\\ University of Durham\\ Durham DH1 3LE, U.K.}\\ 
\end{center}

\begin{abstract}
\noindent A vital constituent of a virus is its protein shell, called the viral capsid, that encapsulates and hence provides protection for the viral genome. Assembly models are developed for viral capsids built from protein building blocks that can assume different local bonding structures in the capsid. This situation occurs, for example, for viruses in the family of Papovaviridae, which are linked to cancer and are hence of particular interest for the health sector. More specifically, the viral capsids of the (pseudo-) $T=7$ particles in this family consist of pentamers that exhibit two different types of bonding structures. While this scenario cannot be described mathematically in terms of Caspar-Klug Theory (Caspar and Klug 1962), it can be modelled via tiling theory (Twarock 2004). The latter is used to encode the local bonding environment of the building blocks in a combinatorial structure, called the assembly tree, which is a basic ingredient in the derivation of assembly models for Papovaviridae along the lines of the equilibrium approach of Zlotnick (Zlotnick 1994). A phase space formalism is introduced to characterize the changes in the assembly pathways and intermediates triggered by the variations in the association energies characterizing the bonds between the building blocks in the capsid. Furthermore, the assembly pathways and concentrations of the statistically dominant assembly intermediates are determined. The example of Simian Virus 40 is discussed in detail. 
\end{abstract}


\section{Introduction}
Papovaviridae are of particular interest for the medical sector because they contain tumour-causing viruses. 
The distinctive feature of viruses in this family is the fact that the surface lattices of the icosahedral viral capsids, that is, the protein shells encapsulating the viral genome, are composed of clusters of five proteins only, called pentamers. 
This structural peculiarity distinguishes them from viruses in other families, and is the reason why the Caspar-Klug theory (Caspar and Klug  1962), which explains the structure of most viruses with overall icosahedral symmetry, cannot be applied to viruses in this family. For examples see  (Rayment et al. 1982), (Liddington et al. 1991) and (Casjens 1985). 

Virus Tiling Theory (VTT) as introduced in (Twarock 2004, 2005) provides mathematical tools appropriate to the description of the capsid structure of Papovaviridae while still reproducing the tessellations (tilings) relevant to the description of the viruses in the Caspar-Klug classification. Its predictive power is significantly enhanced, in comparison with the Caspar-Klug theory, through its ability to locate the {\em bonds} between protein subunits, and not only the location of the protein subunits themselves. The VTT approach both generalises and extends
the Caspar-Klug theory. It is proving a versatile and powerful ally in tackling the puzzles of modern structural virology, one of them being the mechanisms of virus capsid assembly which we investigate in this paper.


Virus capsid assembly has been considered from various points of view in the literature. Besides the approach of molecular dynamics (Rapaport et al. 1999) and of thermodynamics as self-organisation of disks on a sphere (Bruinsma et al. 2003), combinatorial optimisation studies have been performed in (Reddy et al. 1998) and (Horton and Lewis 1992). Closest to our standpoint are the local rules approach (Berger et al. 1994), where assembly is constrained by a set of local rules that indicate possible locally allowed configurations for the protein subunits, and an equilibrium approach due to Zlotnick (Endres and Zlotnick 2002), where kinetic rate equations determine the concentrations of the assembly intermediates. 

A distinctive feature of the Zlotnick approach is the fact that the local bonding structure of the capsomers (regular grouping of capsid proteins) is {\em not} taken into account when constructing the assembly models. While this is an appropriate simplification for the viruses studied by Zlotnick, it does not accurately reflect the fact that the capsids of Papovaviridae are formed from  pentamers that differ by the structure of the inter-subunit bonds that surround them. Here, we take advantage of the tiling approach, which provides mathematical tools for the modelling of the bonding structure of proteins by combining the information on local environments around pentamers with the equilibrium approach of Zlotnick. 
A complication when studying assembly models with several basic types of capsomers (building blocks) and/or several types of bonds along the above lines is the fact that distinct configurations may be energetically preferred when attaching a further building block. The assembly process must therefore be represented as a tree of assembly pathways rather than as a single assembly pathway as in Zlotnick's simplified model. We develop a method to characterize the structure of the assembly trees as a  function of the association energies of the bonds, and show that only a subset of statistically dominant intermediates needs to be taken into account when analysing the assembly process.

The paper is organised as follows. After a review of the tiling model for (pseudo-) $T=7$ capsids in the family of Papovaviridae in Section~\ref{two}, and a review of Zlotnick's assembly model in Section~\ref{three}, we introduce the  assembly models for Papovaviridae in Section~\ref{four}. In particular, we derive the succession of assembly intermediates based on a set of building blocks and rules for their association, and encode this information in an assembly tree.  We use the latter to derive equations for the relative concentrations of the intermediates in solution in an {\em in vitro} experiment. In Section~\ref{five}, we apply this set-up to Simian Virus 40 (SV40) and make predictions about the concentrations of the assembly intermediates which may be tested experimentally. 

\section{Modelling the bonding structure of (pseudo-)\\ $T=7$ capsids in the family of Papovaviridae}
\label{two}

\begin{figure}[ht]
\begin{center}
\includegraphics[width=4cm,keepaspectratio]{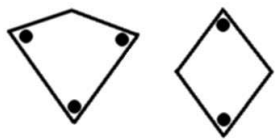}\\*
\end{center}
\caption{\em Tiles for spherical capsids in the family of Papovaviridae.}
\label{fig1}
\end{figure}

While (pseudo-) $T=7$ capsids in the family of Papovaviridae fall out of the orbit of Caspar-Klug Theory, it has been shown in (Twarock  2004) that their surface structure can be modelled via tiling theory. The viral capsids are tessellated in terms of a set of building blocks called tiles and the tilings encode interactions between protein subunits, which are marked schematically as dots on the tiles. In this way, both the locations of the protein subunits and of the inter-subunit bonds can be read off from the tiling. 
The tiles for Papovaviridae are shown in Fig.~\ref{fig1}. 
The tile on the left is called a kite and represents a trimer interaction between the three protein subunits that are represented as dots; the tile on the right is called a rhomb and corresponds to a dimer interaction between the two subunits on the tile. Trimer and dimer interactions take place via an exchange of C-Terminal arms between the subunits. 

\begin{figure}[h]
\begin{center}
\includegraphics[width=5cm,keepaspectratio]{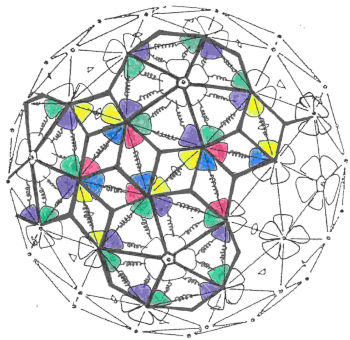}\\
\end{center}
\caption{\em Tiling representing the surface lattice of spherical tilings in the family of Papovaviridae.}
\label{fig2}
\end{figure}

The corresponding tiling is shown in Fig.~\ref{fig2}, superimposed on experimental data from (Liddington et al. 1991). 
 It yields the surface lattice of the (pseudo-) $T=7$ capsids  in the family of Papovaviridae, which are built from 360 protein subunits organised in 72 pentamers.
The dots, which indicate the locations of the protein subunits with respect to the shapes of the tiles,  
are located at angles of size $\theta:= \frac{2\pi}{5}$, so that the corresponding vertices mark the locations of pentamers and the structure of the tiles fixes their relative orientations. 

The locations of the inter-subunit bonds can be read off from the tiling based on the interpretation of the tiles. They have been superimposed schematically as spiral arms on the tiling in Fig.~\ref{fig2} and coincide with the experimentally found bonding structure, for example as observed for papillomavirus in (Modis et al. 2002). 

\begin{figure}[ht]
\begin{center}
\includegraphics[width=6cm,keepaspectratio]{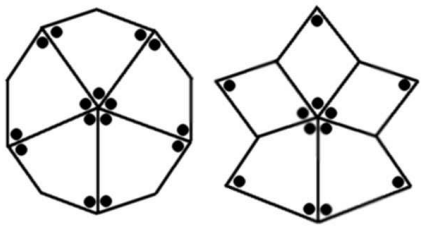}
\end{center}
\caption{\em The two vertex stars in the tiling.}\label{vstars}
\end{figure}

\begin{figure}[ht]
\begin{center}
\includegraphics[width=6cm,keepaspectratio]{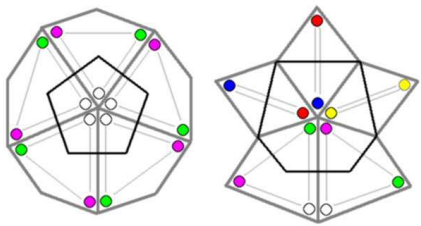}
\end{center}
\caption{\em The building blocks for assembly of vertex-star models.}
\label{V1V2}
\end{figure}

The tiling shows that there are two different types of pentamers in the capsid, distinguished by their local bonding structure, i.e. by the locations and types of the inter-subunit bonds that surround them. This information is encoded in the two different local environments of vertices marking pentamers in the tiling 
(vertex-stars of the tiling), see Fig.~\ref{vstars}. 
In order to avoid overlaps of the building blocks, and indeed to base assembly on the pentamers only rather than pentamers plus surrounding subunits, we choose to work instead with the hexagons and pentagons that are obtained from the building blocks in Fig.~\ref{vstars} by cutting all  bonds perpendicularly through their middle as shown in Fig. 
\ref{V1V2}. 

\section{Zlotnick's model}
\label{three}

In (Zlotnick 1994), viral capsid assembly is described as a series of equilibrium reactions for a small plant virus formed from 12 pentamers. All pentamers  are identical and are modelled geometrically as regular pentagons. This choice of building block is justified by the experimental observation that during assembly pentamers form first and then combine to produce the full capsid. 

\subsection{Assumptions of the model}

For simplicity, the model assumes that the final capsid has ideal dodecahedral geometry and that
assembly takes place by the association of a single subunit at a time to an existing intermediate, rather than by the association of intermediates. 

A crucial hypothesis is that all 30 edge to edge contacts between pentamers (located on the dodecahedral 2-fold axes) are identical and score the same per-contact association energy $\Delta G_{contact}$. This energy is used to define the association constant for a single inter-subunit contact, $K_{contact}$, through the thermodynamical Arrhenius relation $\Delta G_{contact}=-RT \ln K_{contact}$, with $R$ the gas constant and $T$ the temperature in Kelvin, and can therefore be interpreted as a free energy.

The model assumes that ${\Delta}G_{contact} << 0$ kcal mol$^{-1}$, so that the most stable intermediates correspond to those with the greatest number of inter-subunit contacts.



\subsection{The model for the concentrations of the intermediates}

Based on the above assumptions, Zlotnick's model determines the concentrations of the most stable intermediates during assembly. In particular, it determines the assembly pathway for the virus and predicts the relative concentrations of the intermediates on this pathway. Since the pathway is linear in this model (i.e. does not have any branches because any further step is uniquely determined by the assumptions), it is possible to label the intermediates in consecutive order by their location on the pathway. In particular, we denote the $n$-th assembly intermediate (also referred to as species) on the pathway as $(n)$ and its concentration as $[n]$.  

For each iteration step $n$, $K_{n}$ denotes the {\it association constant of the intermediate $(n)$}. It is defined as  
\begin{equation}\label{eq:1}
K_{n} = \frac{[n]}{[n-1][1]}
\end{equation}
and can be decomposed into $K_{n} = S_{d}S_{n}K'_{n}$, where the individual factors have the following interpretation: 

\begin{itemize}
\item Let $O_{sym}(n)$ denote the order of the discrete symmetry group that corresponds to the geometrical shape of intermediate $(n)$. Then 
\begin{equation}\label{eq:2}
S_{n} = \frac{O_{sym}(n-1)}{O_{sym}(n)}.
\end{equation}

\item $S_{d}={O_{sym}(1)}$ corresponds to the order of the discrete rotational symmetry group of the shape of the incoming subunit, i.e. the pentagon. It takes care of the geometrical degeneracy of the incoming subunit. 

\item $K'_{n}$ is a function of the number of contacts formed in the transition from intermediate $(n-1)$ to intermediate $(n)$ and their free energies. It is given as 
\begin{equation}\label{eq:3}
K'_{n} = \exp \left( -\frac{\gamma(n) {\Delta}G_{contact}}{RT}\right)=(K_{contact})^{\gamma(n)},
\end{equation}
where $\gamma(n)$ is the number of inter-subunit bonds formed. 
\end{itemize}

The assembly pathway is shown in Table~\ref{table:ass1}  in Section A1 of the Appendix. 
The concentrations $[n]$ of the assembly intermediates on the pathway can be obtained from Eq.~(\ref{eq:1}) as follows: 
\begin{eqnarray}\label{eq:4}
  [n] &=& S_{d}S_{n}K'_{n}[n-1][1] \nonumber \\
        &=& \left(\prod^{n}_{i=2}S_{d}S_{i}K'_{i}\right)[1]^{n}.
\end{eqnarray}

\subsection{Interpretation of the model} 
\label{33}

Eq.~(\ref{eq:4}) is crucial in Zlotnick's analysis of the concentrations of intermediates. It clearly shows how the concentrations depend on two parameters which can be tuned experimentally at room temperature: one is the free energy of a single contact ${\Delta}G_{contact}$, and the other is the concentration, $[1]$, of basic subunits.

For the choice ${\Delta}G_{contact} << 0$ kcal mol$^{-1}$, one obtains fully disassembled subunits and intact dodecahedra, with rare occurrence of assembly intermediates. This property of the model is illustrated in Table~\ref{table:ass2} by tuning the second parameter, $[1]$, to three significant values.

The first, {\em{pseudo-critical}}, value\footnote{Following (Johnson et al. 2005) and (Tanford 1980) we use the terminology {\em{pseudo-critical}}, because subunits cannot freely equilibrate between the phases.}  is the concentration of basic subunits needed to ensure that the concentration of intact dodecahedra at equilibrium, calculated with Eq.~(\ref{eq:4}) when $n=12$, namely 
\begin{eqnarray}\label{eq:5}
  [12] &=& \left(\prod^{12}_{i=2}S_{d}S_{(i)}K'_{i}\right)[1]^{12} \nonumber \\
       &=& 5^{11}\frac{1}{12}\exp \left(-\frac{30{\Delta}G_{contact}}{RT}\right)[1]^{12},
\end{eqnarray}
satisfies the constraint 
\begin{equation}\label{eq:5a}
        [12]=[1]
\end{equation}
at a fixed value of ${\Delta}G_{contact}$ (negative) and $T$.

Eqs (\ref{eq:5}) and (\ref{eq:5a}) imply that the overall subunit - dodecahedron equilibrium is described by
\begin{equation}\label{eq:5b}
\frac{30{\Delta}G_{contact} - RT\ln\frac{5^{11}}{12}}{11} = RT\ln[1] \equiv {\Delta}G_{app} = -RT\left|\ln \K\right|.
\end{equation}

$\K$ is therefore the dissociation constant for a single contact in the final capsid when the concentration of incoming subunits and final capsid coincide, i.e. one has
\begin{equation}\label{eq:5c}
        \K=[1]=[12].
\end{equation}

The two other chosen values in (Zlotnick 1994) are $[1]={\fr12}\K$ and 
$[1]=2\K$.

The concentrations of $[1]$ in Table~\ref{table:ass2} were chosen to be 
${\fr12}\K$, $\K$ and $2\K$ calculated from Eq.~(\ref{eq:5b}) with ${\Delta}G_{contact} = -2.72$ kcal mol$^{-1}$, $R=1.987$cal K$^{-1}$ mol$^{-1}$ at room temperature $(T=298K)$.

\begin{table}
  \centering
\begin{tabular}{|l|c c c|}
    \hline
    Species & & Concentrations (in molar units M) &  \\ \hline
    1 & $0.44*10^{-6}$ & $0.88*10^{-6}$ & $1.8*10^{-6}$ \\
    2 & $2*10^{-10}$ & $1*10^{-9}$ & $5*10^{-9}$ \\
    3 & $4*10^{-12}$ & $3*10^{-11}$ & $2*10^{-10}$ \\
    4 & $1*10^{-13}$ & $2*10^{-12}$ & $3*10^{-11}$ \\
    5 & $5*10^{-15}$ & $2*10^{-13}$ & $5*10^{-12}$ \\
    6 & $2*10^{-15}$ & $1*10^{-13}$ & $1*10^{-11}$ \\
    7 & $2*10^{-16}$ & $3*10^{-14}$ & $4*10^{-12}$ \\
    8 & $2*10^{-16}$ & $7*10^{-14}$ & $2*10^{-11}$ \\
    9 & $6*10^{-16}$ & $2*10^{-13}$ & $1*10^{-10}$ \\
    10 & $1*10^{-15}$ & $1*10^{-12}$ & $1*10^{-9}$ \\
    11 & $1*10^{-13}$ & $2*10^{-10}$ & $5*10^{-7}$ \\
    12 & $2*10^{-10}$ & $0.88*10^{-6}$ & $3.6*10^{-3}$ \\ \hline
  \end{tabular}
  \caption{\em The concentrations of assembly intermediates at three concentrations of basic subunit [1] (Zlotnick 1994).}\label{table:ass2}
\end{table}

\section{Generalised assembly models and assembly trees}
\label{four}

In this section, a generalisation of Zlotnick's model to multiple  building blocks and types of inter-subunit bonds is introduced. 
It is considered in the context of assembly rules that do not specify a single pathway, but instead lead to an assembly tree that develops several possible pathways. It is shown how the recursion relations for intermediate concentrations can be evaluated on the basis of information encoded in the assembly tree. 

\subsection{Generalisation of Zlotnick's model to N types of building blocks and k types of inter-subunit bonds}

We generalise the equations for the assembly intermediates in Zlotnick's assembly model to the case of $N$ different incoming subunits $1_{i}$, $i=1,...,N$. In this case, the association constant of the intermediate $(n)$ is given by, 
\begin{equation}\label{eq:6}
 K_{n} = \frac{[n]}{[n-1]\sum_{i=1}^{N}[1_{i}]\delta_{i,x(n)}} = d_{x(n)}\,
S_{n}\,K'_{n},
\end{equation}
instead of Eq.~(\ref{eq:1}). In the above expression, $x(n)$ corresponds to the incoming subunit selected among the $N$ possible ones for addition in iteration step $n$, so that $\delta_{i,x(n)}=1$ if $i=x(n)$, and vanishes otherwise. 
This formula thus describes the transition from a specific intermediate $(n-1)$ to a specific intermediate $(n)$ via  attachment of one of the $N$ possible subunits. We have included it for completeness as it is relevant for viral capsids whose building blocks are different in solution (such as CCMV, which is formed from dimers and pentamers of dimers (Johnson et al., 2005)). In contrast, the SV40 virus studied in this paper is formed from building blocks that are identical in solution (pentamers) but are distinguished  by their local environment of bonds once they have formed a capsid.

If $d_i$ denotes the order of the discrete rotational symmetry group associated with the shape of subunit $i$, then $d_{x(n)}$ encodes the geometric degeneracy of the incoming subunit at iteration step $n$. If there are $k$ different types of inter-subunit bonds occurring in the final capsid, each with free energy 
${\Delta}G_j$, ($j=1,\ldots, k$), then one has  
\begin{equation}
K'_{n} = \exp\left( -\frac{\sum_{j=1}^{k} \alpha_j (n)\Delta G_j}{RT}\right)\,,
\end{equation}
where ${\alpha}_j(n)$ denotes the number of bonds of type $j$ formed at step $n$ with free energy ${\Delta}G_j$. 
Note that during the transition from  intermediate $(n-1)$ to intermediate $(n)$  only a subset of the $k$  different types of inter-subunit bonds is formed in most cases, and hence some (but not all) of the ${\alpha}_j(n)$ may be zero. 
As before, $S_{n}$ is defined as the ratio of the order of the discrete rotational symmetry group associated with the geometrical shape of assembly intermediate $(n-1)$ and the order of the symmetry group corresponding to $(n)$. 

Eq.~(\ref{eq:6}) provides an expression for the concentration $[n]$ of intermediate $(n)$, in terms of $[n-1]$, namely, 
\begin{equation}\label{eq:7}
[n] =\Omega(n)[n-1],
\end{equation}
where 
\begin{equation}\label{eq:7a}
\Omega(n)=d_{x(n)}\,S_{n}\,\exp \left(-\frac{\sum_{j=1}^{k} \alpha_{j}(n)\Delta G_j}{RT}\right) \sum_{i=1}^{N}\,[1_{i}]\,{\delta_{i,x(n)}}.
\end{equation}
Formula (\ref{eq:7a}) generalises Eq.~(\ref{eq:4}) in Zlotnick's model.
In order to express $[n]$ in terms of the concentrations of the incoming building blocks, that is $[1_{i}]$, Eq.~(\ref{eq:7}) has to be applied recursively. For this, information on all possible assembly pathways is needed as an input, because it determines the factor $\Omega(n)$ for each iteration step $n$. This information is encoded in the assembly tree as discussed in the following subsection.  

\subsection{Assembly trees}
\label{asstrees}
\begin{figure}[ht]
\begin{center}
\includegraphics[width=10cm,keepaspectratio]{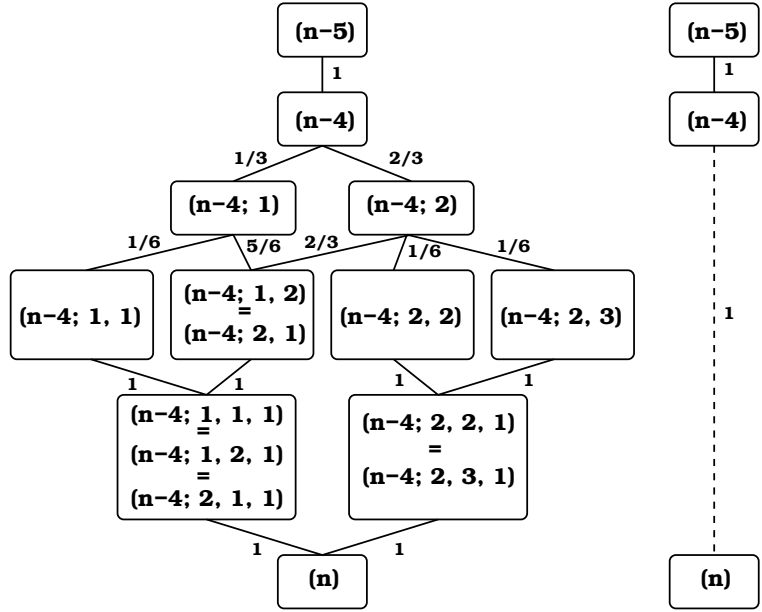}\\*
\end{center}
\caption{\em{Example of primary closed bundle in an assembly tree.}}
\label{Ex:a}
\end{figure}

As already stressed in the introduction,
a common feature of assembly models for Papovaviridae is the occurrence of a number of assembly pathways rather than a single one as in Zlotnick's model of Section~\ref{three}. The collection of assembly pathways and their relations are cast in an assembly tree. Evaluating the law of mass action as in Eq.~(\ref{eq:4}) for such assembly trees is very involved, and it is the purpose of this section to construct associated linear (`branchless') trees (see Fig. \ref{Ex:a}) on which Zlotnick's formula (\ref{eq:4}) can be used. 

We start by  setting up some terminology. 
An assembly tree is a directed graph in which a {\it node} represents an assembly intermediate; a {\it link} between any two nodes is drawn if the intermediates corresponding to these nodes are related to each other by attachment or detachment of a single basic building block. A {\em path} in the assembly tree is a connected subset of nodes and links such that each node has at most one incoming and one outgoing link. A node is called {\em primary} (and the corresponding assembly intermediate a {\it primary intermediate}) if it is located on all paths in the assembly tree. 
A {\em closed bundle} denotes the collection of paths in the assembly tree between two  primary nodes.
It is called {\em primary} if it is located between two consecutive primary nodes. For a concrete example illustrating this terminology, see Fig. \ref{Ex:a}. 


Since primary intermediates, by definition, appear on each path in the assembly tree, they occur with a higher frequency than the other intermediates. Therefore, we focus here on the concentrations of the primary intermediates only. 

Let $(n-s)$ and $(n)$ be two consecutive primary nodes in the assembly tree. Then the factor ${\tilde \Omega}(n,s)$ with 
$[n]={\tilde \Omega}(n,s) [n-s]$ is called the {\it transition factor} of the primary bundle between these nodes. 
It is the aim in the remainder of this section to derive a closed formula for the transition factor between any consecutive primary nodes in an assembly tree. 

We start with the following result. 
\begin{lemma}\label{lem1}
Along each path in a closed primary bundle, the same number of each type of incoming subunit is added, and the same number of inter-subunit edge to edge bonds is formed.
\end{lemma}

\begin{proof}
All of the paths in a closed primary bundle originate from, and terminate at, a primary node. Since these nodes have a well defined, unique structure in terms of the number of subunits and their location, the same number and types of subunits have been added on each path in the bundle, and the same number of inter-subunit bonds have been formed.
\end{proof}

This implies that there are the same number of nodes on each path in a closed primary bundle. Therefore, one can consider without loss of generality the case of closed primary bundles consisting of $p$ paths with an equal number of nodes each. 

In order to formulate a recursion relation for the transition between two consecutive primary nodes $(n-s)$ and $(n)$, one must introduce a labelling system that uniquely specifies each assembly intermediate according to the path chosen from the primary node $(n-s)$ to form it.
We represent each path between $(n-s)$ and $(n)$ as a unique $s$-dimensional vector $(i_j), 1 \le j \le s $, whose $j$-th component specifies the branch taken at the intermediate node $(n-s+j-1)$. For any value of $s$ and $j$ in the range $1 \le j \le s $, let $(n-s;\,i_1,...,i_j)$ denote the assembly intermediate obtained from $(n-s)$ after addition of $j$ building blocks along the path labelled by the indices $i_1,...,i_j$. These encode the branchings chosen between the primary node $(n-s)$ and the node $(n-s+j)$. 

If $b((n-s))$ (resp. $b((n-s;\,i_1,..,i_j))$\,) denotes the number of outgoing branches at node $(n-s)$ (resp. at node $(n-s;\,i_1,...,i_j)$\,), and if $p_{i_1}((n-s))$ (resp. $p_{i_j}((n-s;\,i_1,...,i_j)), 2 \le j \le s$) is the probability that the branch labelled $i_1$ (resp. $i_j$) is chosen at node $(n-s)$ (resp. $(n-s;\,i_1,...,i_{j-1})$\,), the announced recursion relation for the transition between two consecutive primary nodes $(n-s)$ and $(n)$ is given by,
\begin{eqnarray}\label{eq:15}
 &&   [n]=\left  \lbrace \sum^{b((n-s))}_{i_{1}=1}\,\sum^{b((n-s;\,i_{1}))}_{i_{2}=1}\ldots \sum^{b((n-s;\,i_{1},\ldots,i_{s-1}))}_{i_{s}=1} \, p_{i_1}((n-s))\,\Omega(n-s+1)_{i_{1}} \right.\nonumber\\
    &&\left. \phantom{\sum_{i_1}^b}\times p_{i_2}((n-s;\,i_{1}))\,\Omega(n-s+2)_{i_{2}}\times \ldots \times p_{i_{s}}((n-s;\,i_{1},\ldots,i_{s-1}))\, \Omega(n)_{i_{s}}\right \rbrace \nonumber\\ 
    &&\mbox{}\hskip .6cm \times [n-s],
\end{eqnarray}
with 
\begin{equation}\label{norm1}
    \sum^{b((n-s))}_{i_1=1} p_{i_1}((n-s))=1 
\end{equation}
and
\begin{equation}\label{norm2}
    \sum^{b((n-s;\,i_{1},\ldots,i_{j-1}))}_{i_{j}=1}p_{i_{j}}((n-s;\,i_{1},\ldots,i_{j-1}))=1\,,\,\,  j\in\lbrace 2, \ldots, s\rbrace\,. 
\end{equation}

Note that the factors $\Omega(n-s+j)_{i_{j}}$, $1\le j\le s$  depend on the geometric degeneracy of the incoming subunit and on the association energy of each bond formed at each step $j$, the symmetry of the intermediate $(n-s;\,i_1,\ldots i_j)$,  and the concentrations of the basic subunits. The following example illustrates the general formalism just introduced.

Fig. \ref{Ex:a} shows a primary closed bundle in an assembly tree, i.e. of the collection of assembly pathways between two consecutive primary nodes $(n-4)$ and $(n)$.
With respect to the notations used in (\ref{eq:15}) this corresponds to the situation where $s=4$.  The number of branches stemming from the node $(n-4)$ is 
$b((n-4))=2$, hence $i_1\in\lbrace 1,2\rbrace$.  Moreover,  $b((n-4;\,1))=2$ and $b((n-4;\,2))=3$. The probabilities marked on the links are chosen such that the relations (\ref{norm1}) and (\ref{norm2}) are fulfilled; any other choice that satisfies these equations would also have been possible. Here, one has 
$p_1((n-4))=1/3$, $p_2((n-4))=2/3$, $p_1((n-4;\,1))=1/6$, $p_2((n-4;\,1))=5/6$, $p_1((n-4;\,2))=2/3$,
$p_2((n-4;\,2))=p_3((n-4;\,2))=1/6$ etc. Theorem \ref{thm1} below shows that, from the point of view of computing the concentration of the assembly intermediate $(n)$ from the assembly intermediate $(n-4)$ via the law of mass action, the result is the same as using the reduced tree (single pathway) shown on the right in Fig. \ref{Ex:a}, where the dashed line corresponds to any of the pathways in the bundle without statistical factor correction. 

Let us emphasize that in principle the concentrations of all intermediates can be calculated using the law of mass action, but corrective factors based on the probabilities in Eqs (\ref{norm1}) and (\ref{norm2}) appear. 
For example, the assembly intermediate denoted as $(n-4;\,2,2)$ in  Fig. \ref{Ex:a} has probability
$p_2((n-4)) \times p_2((n-4;\,2))=2/3 \times 1/6=1/9$ of being formed from the primary node 
$(n-4)$.
In contrast to this, no corrective factors appear for primary intermediates as we show in the theorem below. 

\begin{theorem}\label{thm1}
For each closed primary bundle between two consecutive primary nodes $(n-s)$ and $(n)$, the concentration $[n]$ of the primary intermediate $(n)$ is given in 
terms of the concentration $[n-s]$ of the primary intermediate $(n-s)$ by the following equation,
\begin{eqnarray}\label{eq:18}
    [n] &=& {\Omega(n-s+1)}{\Omega(n-s+2)}\ldots {\Omega(n-1)}{\Omega(n)}[n-s] \nonumber \\
    &=& \left(\prod^{s-1}_{j=0}\Omega(n-j)\right)[n-s]
\end{eqnarray}
with $\Omega(n-j)$ as in Eq.~(\ref{eq:7a}).
\end{theorem}

\begin{proof}
Due to Lemma~\ref{lem1}, the same subunits are added on each path in a closed primary bundle (though possibly in a different order). Hence, the transition factor must be proportional to $\left(\prod^{s-1}_{j=0}\Omega(n-j)\right)$, i.e. the transition factor introduced at the beginning of the present section is  
\begin{equation}\label{eq:tf}
    {\tilde \Omega(n,s)} =  C \left(\prod^{s-1}_{j=0}\Omega(n-j)\right)
\end{equation}
with a constant $C$. 
Moreover, due to the normalisation of the probabilities, one has $C=1$. 
\end{proof}

Note that Eq.~(\ref{eq:18}) resembles Eq.~(\ref{eq:2}) in Zlotnick's model. However, while Zlotnick's model considers all intermediates, this formula is defined for concentrations of {\em primary} intermediates only. This choice is justified  by the fact that primary intermediates are statistically dominant because they appear on all paths in the assembly tree. 

Using the recursion relation of Theorem~\ref{thm1}, the concentration of the primary intermediate $(n)$ can be expressed in terms of the concentrations of the incoming subunits using Eq.~(\ref{eq:18}). The formula depends on the number of incoming subunits $1_{1},...,1_{N}$, their geometrical degeneracies $d_1,...,
d_N$ and the total number of inter-subunit bonds $k$ with free energies 
$\Delta G_1,...,\Delta G_k$. One obtains
\begin{equation}\label{eq:19}
    [n] = \left(\prod^{n}_{j=1}\Omega(j)\right)\,, 
\end{equation}
with $\Omega(j)$ as in Eq.~(\ref{eq:7a}).
In particular, for the concentration of the final capsid, $[\mbox{ Final Capsid }]$, formed after $\nu$ steps and after $\alpha_j(\nu)$ bonds of type $j$ being added, one obtains: 
\begin{equation}\label{eq:22}
    [\mbox{ Final Capsid }] = \left(\prod^{N}_{i=1}(d_i\,[1_i])^{\eta(i)}\right)\frac{1}{s}\exp\left(-\frac{\sum^{k}_{j=1}\alpha_j(\nu)\Delta G_{j}}{RT}\right).
\end{equation}
Here $\eta(i)$ is the number of subunits $1_i$ in the final capsid and  $s$ is the order of the discrete rotational symmetry group of the final capsid.

\section{Assembly models for Papovaviridae}
\label{five}

Here the set-up of Section~\ref{four} is applied to (pseudo-) $T=7$ viruses in the family of Papovaviridae. 
As discussed in Section~\ref{two}, these viruses have capsids formed from 360 proteins which organise themselves in two different types of pentamers that are distinguished by their local bonding environments. These, together with the locations and relative orientations of the pentamers on the capsid, can be modelled via tiling theory. 
As can be seen from the tiling in Fig.~\ref{fig2}, there are two different types of basic shapes, a rhomb and a kite, each representing a different type of inter-subunit bond. Hence, the tiling approach provides a natural way to  model different types of local environments mathematically: the pentamers located at the twelve global five-fold axes of the tiling  are surrounded by five kite tiles, and the sixty other pentamers are surrounded by three rhomb tiles and two kite tiles each.  

The two different types of pentamers are modelled as two different vertex configurations, called vertex-stars in tiling theory
(see Fig.~\ref{vstars}).  The corresponding model is called the {\it vertex-star model} and is discussed in Subsection~\ref{fiveone}.

In Subsection~\ref{fivetwo}, we apply the vertex-star model to SV40 and use it to compute the concentrations of the primary assembly intermediates.

\subsection{The vertex-star model}
\label{fiveone}

This model is based on the assumption that all pentamers form first in solution as supported by experimental evidence (Kanesashi et al. 2003) and then assemble to build the capsid. Since the pentamers acquire their diversity only after they are bound in the caspid, we do not distinguish between the two types of building blocks of Fig.~\ref{V1V2} in solution. The vertex star model generalizes Zlotnick's model in two ways. Although there is only one type ($N=1$) of incoming subunits, namely the pentamers, there are $k=3$ types of intersubunit bonds {\em and} different local bonding environments for the pentamers. 

We now describe these environments in some detail.
While the bonds on the five edges of the pentagonal building block associated with the pentamer on the right in Fig.~\ref{V1V2} are all identical, they differ for the hexagonal building block associated with the pentamer on the left in that figure. The corresponding association energies are labelled as follows: 
\begin{itemize}
\item We denote the association energy corresponding to a single C-terminal arm in a trimer (represented by a kite in the tiling model) as $a$. As such, all five edges of the pentagonal building block (corresponding to the pentamers at the 5-fold axes) have an association energy of $2a$. It occurs once on the hexagonal shape, with its neighbouring bonds having an association constant $a$, corresponding to a single C-terminal arm each. 
\item $b$ labels the association  energy related to quasi-dimer bonds. These are located around the 3-fold axes of the tiling and are shown as rhombs with red and blue decorations in Fig.~\ref{fig2}. 
\item $c$ labels the association energy corresponding to strict dimer bonds along the global 2-fold axes of the tiling. These correspond to the rhombs with yellow decorations in Fig.~\ref{fig2}.
\end{itemize}

Note that the distinction between the association energies for bonds on local and global 2-fold axes is important as they are known to be different experimentally, see e.g. (Salunke 1986), (Schwartz 2000). 

\begin{figure}[ht]
\begin{center}
\includegraphics[width=8.5cm,keepaspectratio]{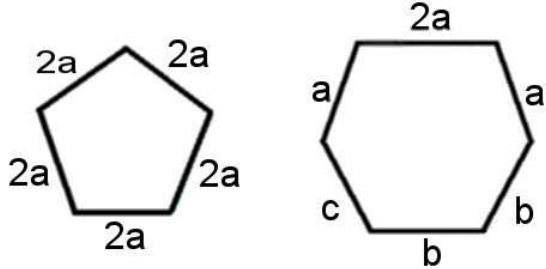}\\*
\end{center}
\caption{\em{Local bonding environments for the pentagonal and hexagonal building blocks.}}
\label{fig6new}
\end{figure}

We can now revisit Fig.~\ref{V1V2} and quantify, in terms of $a,b$ and $c$, the association energies related to all edge-to-edge contacts of the pentagonal and hexagonal building blocks, as shown in Fig.~\ref{fig6new}.

We consider the association energies $a,b$ and $c$ as parameters and start by giving all possible increases in free energy through the addition of a new building block to an existing assembly intermediate. Clearly, the more contacts the incoming building block makes with the latter, the larger the free energy. Up to six bonds may be formed in a single step, and the exhaustive list of possible association energies relative to these bonds is as follows:


\begin{itemize}
\item For single contacts: $a, 2a, b, c$  

\item For two contacts: $3a, a+b, 2b, b+c, a+c, 4a, 2a+b, 2a+c$

\item For three contacts: $3a+b, a+2b, 2b+c, a+b+c, 3a+c, 6a, 2a+b, 2a+c, \break 2a+b+c, 2a+2b$

\item For four contacts: $3a+2b, a+2b+c, 4a+c, 4a+b, 8a, 2a+b+c, 2a+2b+c, 2a+2b$

\item For five contacts: $3a+2b+c, 2a+2b+c, 4a+b+c, 4a+2b, 10a$

\item For six contacts: $4a+2b+c$

\end{itemize}
We assume that the guiding principle for assembly is the maximisation of the free energy of the assembly intermediates at each step. In other words, assembly takes place in such a way that the total free energy associated with the newly formed bonds is maximised. Such optimization determines the structure of the assembly tree, which therefore depends on the relative values of the association energies $a,b$ and $c$. Our aim is to depict the dependence of the assembly tree structure on the ratios $a/c$ and $b/c$ via a 2-dimensional { \em phase space} whose coordinate axes represent $a/c$ and $b/c$ respectively, and where every point represents the assembly tree corresponding to a particular choice of those ratios of association energies.

\begin{figure}[ht]
\begin{center}
\includegraphics[width=5cm,keepaspectratio]{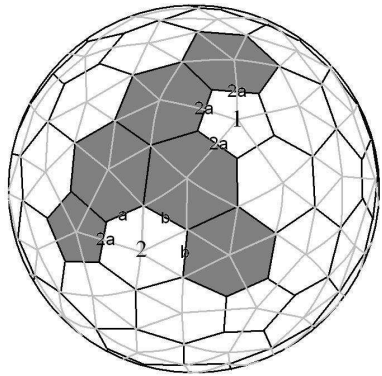}\\
\end{center}
\caption{\em{An assembly intermediate shown in grey with potential next sites labelled by letters.}}
\label{Fign1}
\end{figure}

The idea is to partition the phase space in disjoint domains such that points in a given domain represent identical assembly trees, i.e. have the same assembly intermediates organised on the same assembly pathways. 
By doing so, one can read off from the phase space representation what changes in $a/c$ and $b/c$ affect the structure of a given assembly tree. For instance, if a single contact is realised at step $n$, the tree might be different if, say, $a>b$ rather than $a<b$, in which case one must further discuss whether $b > 2a$ or $b <2a$. The threshold for potential qualitative changes in the assembly tree in this example are therefore the equalities $a=b$ and $b=2a$. 
We demonstrate the geometrical meaning of this in Fig. \ref{Fign1}. 
In this case, the insertion of a building block at position $1$ corresponds to the formation of three bonds of association energy $2a$ each (total energy of $6a$) and the insertion of a building block at position $2$ corresponds to the  formation of four bonds of association energies $2a, a, b, b$ (total energy of $3a+2b$). The relative values of $a$ and $b$  determine which of the two positions will be selected in order to form the next assembly intermediate on the pathway. 

In general, the partition of the phase space is provided by a grid of lines whose equations are obtained by relating pairwise the free energies listed above, namely,
\begin{itemize}
\item Equations relating $a$ and $b$:

$a=b, a=2b, a=b/3, a=b/4, a=b/8, a=b/10, a=b/6, a=b/2, a=2b/3, \break a=b/5,
a=b/7, a=b/9, a=2b/5, a=2b/7, a=2b/9$

\item Equations relating $a$ and $c$:

$a=c, a=c/3, a=c/4, a=c/6, a=c/8, a=c/10, a=c/2, a=c/5, a=c/7, \break a=c/9$

\item Equations relating $b$ and $c$:

$b=c, b=c/2$

\item Mixed cases:

$a=(b+c)/2, a=(2b+c)/2, a=b+c, a=2b+c, a=b-c, a=(b-c)/3, \break a=(b-c)/4,
a=c-b, a=(c-2b)/4, a=(b+c)/3, a=(2b+c)/3, a=(b-c)/2, \break a=2b-c, a=(2b-c)/3,
a=(2b-c)/4, a=(b+c)/4, a=(b+c)/6, a=(b+c)/8, \break a=(b+c)/10, a=(c-b)/2,
a=(c-2b)/2, a=(2b+c)/4, a=(2b-c)/2, \break a=(2b+c)/6, a=(2b+c)/8, a=(2b+c)/10,
a=(b+c)/5, a=(b+c)/7, \break a=(b+c)/9, a=(2b+c)/5, a=(2b+c)/7, a=(2b+c)/9$
\end{itemize}

\begin{figure}[ht]
\begin{center}
\includegraphics[width=13cm,keepaspectratio]{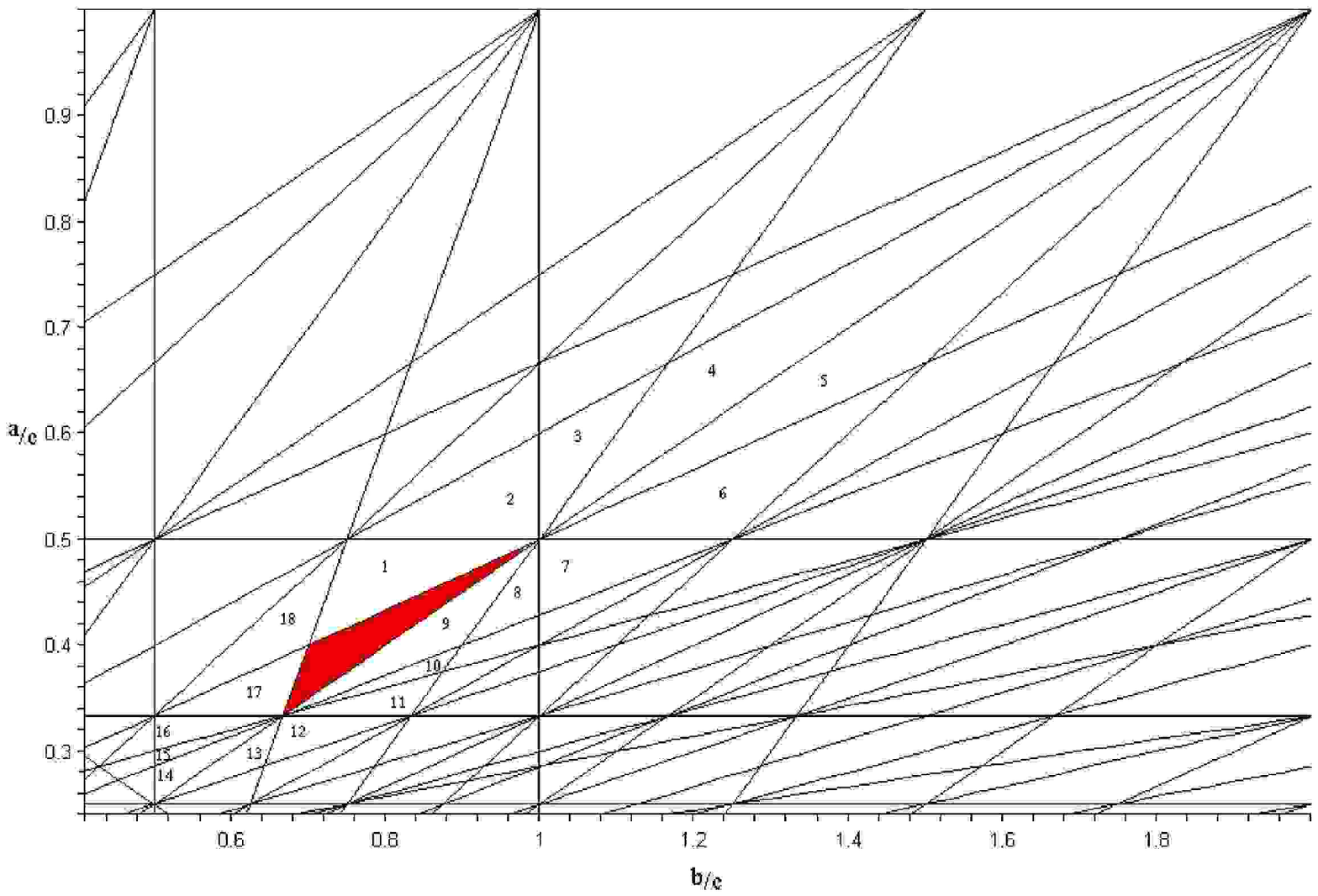}
\end{center}
\caption{\em The phase space formed by $a/c$ and  $b/c$ partitioned into areas of equal primary intermediates.}
\label{Phase}
\end{figure}

In Fig.~\ref{Phase} we show a region of the phase space for the case where $c$ is normalised to 1 and plot  $y \equiv a/c$ (vertical axis) against $x \equiv b/c$ (horizontal axis). 
Note that several domains of the phase space presented in Fig.~\ref{Phase} may exhibit the same assembly trees  because the lines only mark {\em potential} changes. It follows that in practice, the effective phase space has a `lower resolution' than that of Fig.~\ref{Phase}. 

In order to determine the location of the assembly tree of a given virus in phase space, we need to compute $a/c$ and $b/c$. For this, we use the {\em ratios} of the association energies listed on the VIPER webpages, which yield, for SV40, $x=0.92$ and $y=0.47$. Therefore, SV40 lies in the area bounded by the following lines:

\begin{equation}\label{simian}
y=\frac{x}{2}, \quad y=2x-1 \quad {\rm and}\qquad y=\frac{x}{3}+\frac{1}{6}\,.
\end{equation}

In Section A3 of the Appendix, we discuss the qualitative behaviour of all intermediates located in areas of phase space adjacent to SV40. This information can be used to understand how enforcing quantitative changes to the association energies of SV40 can lead to qualitatively different behaviours. 

In Section~\ref{fivetwo} we discuss the case of SV40 in detail, and compute the concentrations of the primary assembly intermediates. 

\subsection{Application to SV40}
\label{fivetwo}

\begin{figure}[ht]
\begin{center}
\includegraphics[width=8cm,keepaspectratio]{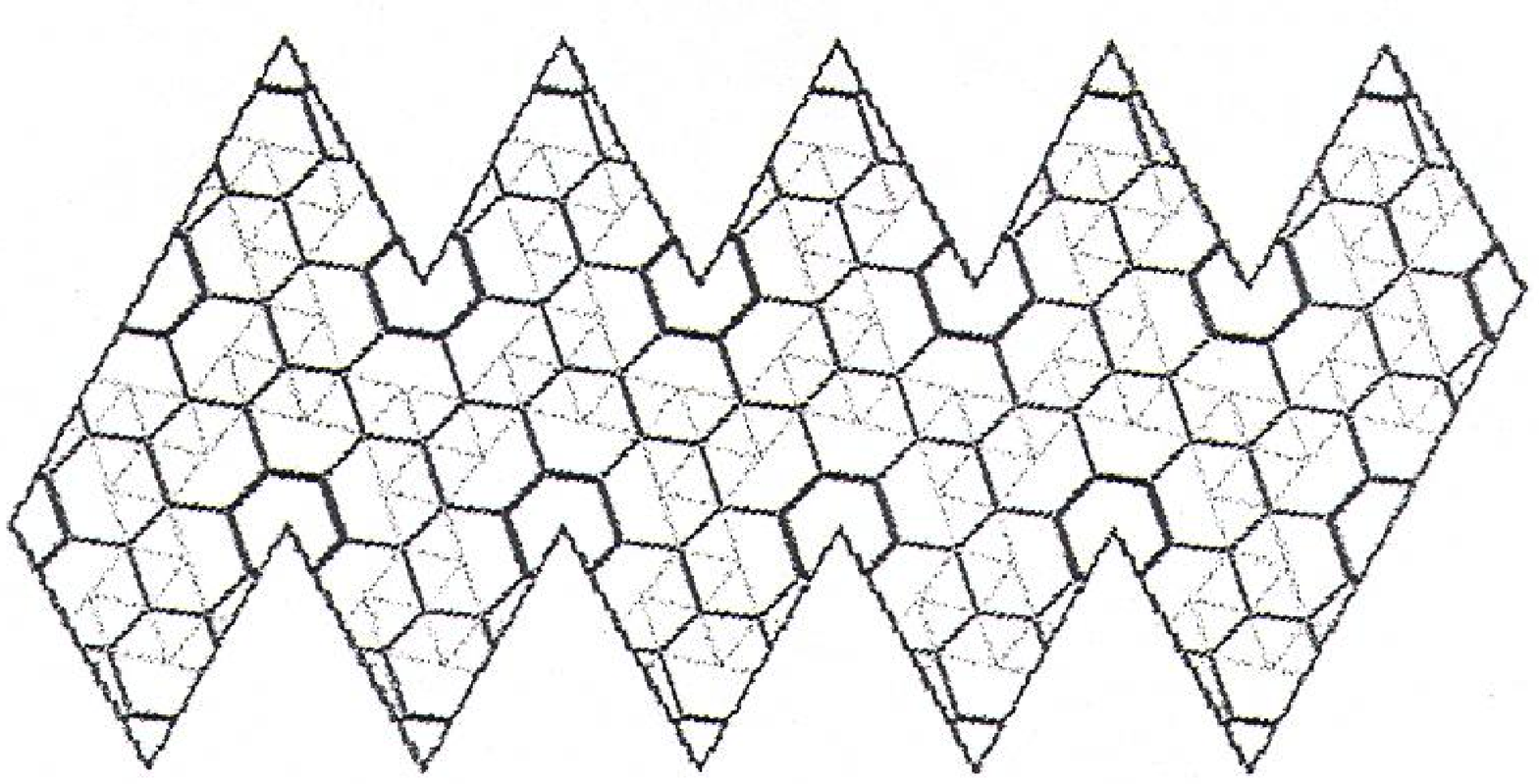}
\end{center}
\caption{\em Template for assembly of vertex-star models.}
\label{V1V2Template}
\end{figure}

In this section we apply the vertex-star model to the case of SV40.
In order to compute the concentrations of the assembly intermediates based on Eq.~(\ref{eq:19}), one needs information on the assembly tree and hence on a generalisation of Table~\ref{table:ass2} to the primary intermediates in our model. 
Embedding the complete capsid into the plane provides an outline of locations of the tiles in the final capsid as shown in Fig.~\ref{V1V2Template}. Capsid assembly can hence be displayed graphically by filling the blank spaces.
In Table~\ref{SV40table1}, Section A3 of the Appendix, where all assembly intermediates are listed for SV40, this will be used in order to display the structure of the primary assembly intermediates graphically. 

Based on this, we compute the concentration of the final capsid in terms of the concencentration of the incoming subunits. Recall that pentamers are assumed to be indistinguishable in solution and correspond to clusters of five proteins, with five C-terminal arms that are not in a particular configuration, but attain a configuration corresponding to one of the two vertex stars only after attachment to the capsid. Therefore, we denote the concentration of the indistinguishable pentamers in solution as [1], and use the fact that their geometrical degeneracy is 5. 
Hence  
\begin{equation}\label{eq:20}
    [\mbox{ Final Capsid }] = (5[1])^{72}\frac{1}{60}\exp \left (-\frac{180a+60b+30c}{RT}\right )\,.
\end{equation} 
The factor 60 in the denominator corresponds to the order of the symmetry group of the final capsid as can be seen by comparison with Eq.~(\ref{eq:2}), noting that 
\begin{equation}
\prod_{n=2}^{72} (S_x S_{n}) = \frac{5^{71} O_{sym}(1)}{O_{sym}(72)} = \frac{5^{72}}{60}\,.
\end{equation}

The factors $\Omega(P)$ for the primary intermediates can be obtained as follows: 
\begin{equation}
\Omega(P) = (5[1])^{\epsilon(P)} \exp \left(-\frac{\alpha(P) a+ \beta(P) b + \gamma(P) c}{RT}\right )\,,
\end{equation}
where $\epsilon(P)$ corresponds to the number of building blocks added in transition from primary assembly intermediate $P-1$ to primary assembly intermediate $P$, and $\alpha(P)$, respectively $\beta(P)$ and $\gamma(P)$, correspond to the number of $a$, respectively $b$ and $c$ bonds, added during that iteration step. 
The corresponding values for SV40 can be read off from Table~\ref{table:polytable1}, Table~\ref{table:polytable2} and Table~\ref{table:polytable3}, 
Section A2 
of the Appendix. 

We compute the concentrations of the assembly intermediates based on starting values given by the dissociation constants as in Subsection~\ref{33}. However, since there are different types of bonds in this model, we use the average dissociation constant here, which will also be denoted as $\K$. In Table~\ref{concentrations} we give the concentrations of the primary assembly intermediates for SV40 for the starting values $\frac12 \K$, $\K$ and $2\K$, where $\K$ is calculated at room temperature with free energies 
\footnote{In contrast to the ratios $x$ and $y$ relevant to the analysis of the phase space in Section 5.1, the absolute values of the association energies on the VIPER webpages are higher than the ones usually observed in experiments. This discrepancy may be explained by the fact that the data on the VIPER webpages have been obtained via computations based on the scenario of rigid building blocks, which may be too strong an assumption (Zlotnick, private communication). Therefore we are using the range of values suggested in (Keef et al. 2005) together with the relative ratios stemming from the VIPER data.}\,$a=-0.7$\,kcal\,mol$^{-1}$,  $b=-1.37$\,kcal\,mol$^{-1}$ and $c=-1.49$\,kcal\,mol$^{-1}$. 
\begin{table}
\centering
\begin{tabular}{|c|c c c|}
    \hline
    Species & \multicolumn{3}{|c|}{Concentrations (in molar units M)} \\ \hline
    1 & $1.822*10^{-4}$ & $3.643*10^{-4}$ & $7.287*10^{-4}$ \\ \hline
    2 & $6.751*10^{-6}$ & $2.700*10^{-5}$ & $1.080*10^{-4}$ \\ \hline
    3 & $5.212*10^{-7}$ & $4.170*10^{-6}$ & $3.336*10^{-5}$ \\ \hline
    4 & $5.372*10^{-8}$ & $8.594*10^{-7}$ & $1.375*10^{-5}$ \\ \hline
    5 & $8.261*10^{-9}$ & $2.644*10^{-7}$ & $8.459*10^{-6}$ \\ \hline
    8 & $1.690*10^{-11}$ & $4.327*10^{-9}$ & $1.108*10^{-6}$ \\ \hline
    9 & $2.600*10^{-12}$ & $1.331*10^{-9}$ & $6.815*10^{-7}$ \\ \hline
    10 & $6.685*10^{-13}$ & $6.845*10^{-10}$ & $7.010*10^{-7}$ \\ \hline
    11 & $6.464*10^{-14}$ & $1.324*10^{-10}$ & $2.711*10^{-7}$ \\ \hline
    12 & $2.654*10^{-14}$ & $1.087*10^{-10}$ & $4.453*10^{-7}$ \\ \hline
    68 & $1.282*10^{-28}$ & $3.783*10^{-8}$ & $1.117*10^{+13}$ \\ \hline
    69 & $1.405*10^{-28}$ & $8.295*10^{-8}$ & $4.897*10^{+13}$ \\ \hline
    70 & $4.696*10^{-28}$ & $5.544*10^{-7}$ & $6.545*10^{+14}$ \\ \hline
    71 & $1.634*10^{-26}$ & $3.859*10^{-5}$ & $9.111*10^{+16}$ \\ \hline
    72 & $7.715*10^{-26}$ & $3.643*10^{-4}$ & $1.720*10^{+18}$ \\ \hline
  \end{tabular}
\caption{\em Concentrations of species at primary nodes based on $\frac12 \K$, 
$\K$ and $2\K$ respectively.}
\label{concentrations}
\end{table}

As in Zlotnick's case we observe that pentameric subunits and final capsids are the dominant species. 

\section{Conclusion and outlook}

We have extended Zlotnick's assembly model to the scenario where the building blocks (capsomers) assume different local bonding configurations in the capsid. This has led to the occurrence of assembly trees rather than a linear pathway of assembly.  We have characterised the structural dependence of these assembly trees on the ratios of the association energies via phase space portraits (Subsection \ref{fiveone}). This approach has allowed us to determine the structure of the 
statistically dominant intermediates, the primary nodes introduced in Subsection \ref{asstrees}. 
A formula has been derived (Eq.~\ref{norm2}), which provides a mean to calculate concentrations
of primary assembly intermediates as a function of the concentration of their primary predecessor in the assembly pathway. Although it is unnecessarily complicated as long as one is interested in obtaining the transition factors between primary intermediates (the latter are independent of the path chosen as shown in Theorem \ref{thm1}), the formula encodes statistical factors relevant for the calculation of concentrations of assembly intermediates {\em within} a bundle as demonstrated for the example in Fig. \ref{Ex:a}.

While the set-up discussed here lends itself, by construction, to an equilibrium analysis of the assembly process, it can  provide clues only on possible assembly kinetics. This issue is discussed in detail in (Keef et al. 2005) where the probability distribution of all intermediates is computed explicitly via a master equation approach and is used to obtain information on the putative pathways of assembly. 

The theory presented here opens up various avenues for applications. For example, it provides a method for determining experimentally the values of the association constants by measuring the equilibrium concentrations of the basic building blocks (pentamers). In particular, by choosing $[\mbox{ Final Capsid }]=[1]$ as in (Zlotnick 1994), (\ref{eq:20}) yields the relation 
\begin{equation}
{{[1]}_{eq}} = \left( \frac{12}{5^{71}} \exp \left( \frac{\kappa a}{RT}\right)\right)^{\frac{1}{71}}\,,
\end{equation}
where $\kappa = 180 + 60 \frac{b}{a} + 30 \frac{c}{a}$ is given in
terms of the ratios of the association energies b and a, as well as c and a.  As before, assuming that the relative values (ratios) of the association energies given on the VIPER webpage are a good approximation (while their absolute values are too large), as discussed earlier, these can be used  to obtain $\kappa \approx 360.8571$. Hence the association energy $a$ is given by 
\begin{equation}
a({{[1]}_{eq}})= \frac{RT}{\kappa}\ln\left( \frac{(5{{[1]}_{eq}})^{71}}{12} \right)\,
\end{equation}
and can be determined via an experimental measurement of the equilibrium concentrations ${{[1]}_{eq}}$ of the pentamers. 

Moreover, the phase space analysis introduced here provides a tool to control changes in the structure of assembly trees, and hence of assembly intermediates and pathways of assembly, as a function of variations in the association energies. It can also be potentially used in conjunction with a more general set-up, where different types of capsids are obtained from the protein building blocks after engineering of changes in their polypeptide chain (see for example (Johnson 2005)). In such a setting, the phase space analysis would give a quantitative prediction on the  association energies needed to trigger the desired outcome, and would hence provide guidance when triggering changes in the polypeptide chains. Our approach could therefore lead to applications in the engineering of nanoparticles and nanocontainers.

\section*{Acknowledgements}
RT has been supported by an EPSRC Advanced Research Fellowship. TK has been supported by the EPSRC grant GR/T26979/01. 


\section*{Bibliography}


\noindent
Zlotnick A.
\newblock To build a virus capsid: An equilibrium model of the self assembly of
  polyhedral protein complexes.
\newblock {\em J. Mol. Biol.}, 241(1):59, 1994.
\smallskip

\noindent
B.~Berger {\em et al}.
\newblock Local rule mechanism for selecting icosahedral shell geometry.
\newblock {\em Discrete Applied Mathematics}, 104(1):97, 2000.
\smallskip

\noindent
B.~Berger {\em et al}.
\newblock A local rule based theory of virus shell assembly.
\newblock {\em Proc. Natl. Acad. Sci.}, 91(16):7732, 1994.
\smallskip

\noindent
R.F. Bruinsma {\em et al}.
\newblock Viral self-assembly as a thermodynamic process.
\newblock {\em Physical Review Letters}, 90:24, 2003.
\smallskip

\noindent
S.~Casjens, editor.
\newblock {\em Virus Structure and Assembly}.
\newblock Jones \& Bartlett, Boston, Mass., 1985.
\smallskip

\noindent
D.~L.~D. Caspar and A.~Klug.
\newblock Physical principles in the construction of regular viruses.
\newblock {\em Cold Spring Harbor Symp. Quant. Biol.}, 27:1, 1962.
\smallskip

\noindent
D.~Endres and A.~Zlotnick.
\newblock Model-based analysis of assembly kinetics for virus capsids or other
  spherical polymers.
\newblock {\em Biophysical Journal}, 83:1217, 2002.
\smallskip

\noindent
N.~Horton and M.~Lewis.
\newblock Calculation of the free energy of association for protein complexes.
\newblock {\em Protein Sci.}, 1:169, 1992.
\smallskip

\noindent
J.~M. Johnson {\em et al}.
\newblock Regulating self-assembly of spherical oligomers.
\newblock {\em Nano Letters}, 5:765, 2005.
\smallskip

\noindent
S.~N. Kanesashi {\em et al}.
\newblock Simian virus 40 vp1 capsid protein forms polymorphic assemblies in
  vitro.
\newblock {\em J. Gen. Virol.}, 84:1899, 2003.
\smallskip

\noindent
T.~Keef.
\newblock An equilibrium assembly model applied to Murine Polyomavirus.
\newblock {\em J. Theor. Med.}, 6(2):91, 2005.
\smallskip

\noindent
R.~C. Liddington {\em et al}.
\newblock Structure of simian virus 40 at 3.8-$\dot{A}$ resolution.
\newblock {\em Nature}, 354:278, 1991.
\smallskip

\noindent
Y.~Modis, B.~L. Trus, and S.~C. Harrison.
\newblock Atomic model of the papillomavirus capsid.
\newblock {\em The EMBO Journal}, 21(18):4754, 2002.
\smallskip

\noindent
Salunke~D. M., Caspar D.~L. D., and Garcea~R. L.
\newblock Self-assembly of purified polyomavirus capsid protein vp1.
\newblock {\em Cell}, 46(6):895, 1986.
\smallskip

\noindent
D.~Rapaport, J.~E. Johnson, and J.~Skolnick.
\newblock Supramolecular self-assembly: Molecular dynamics modeling of
  polyhedral shell formation.
\newblock {\em Computer Physics Communications}, 121:231, 1999.
\smallskip

\noindent
I.~Rayment {\em et al}.
\newblock Polyoma virus capsid structure at 22.5 å resolution.
\newblock {\em Nature}, 295:110, 1982.
\smallskip

\noindent
V.~S. Reddy {\em et al}.
\newblock Energetics of quasiequivalence: Computational analysis of
  protein-protein interactions in icosahedral viruses.
\newblock {\em Biophys. J.}, 74(1):546, 1998.
\smallskip

\noindent
V.~S. Reddy {\em et al}.
\newblock Virus particle explorer (viper), a website for virus capsid
  structures and their computational analyses.
\newblock {\em J. Virol.}, 75(24):11943, 2001.
\smallskip

\noindent
R.~S. Schwartz, R.~L. Garcea, and B.~Berger.
\newblock `local rules' theory applied to polyomavirus polymorphic capsid
  assemblies.
\newblock {\em Virology}, 268(2):461, 2000.
\smallskip

\noindent
C.~Tanford.
\newblock {\em The Hydrophobic Effect: Formation of Micelles and Biological
  Membranes}.
\newblock John Wiley and Sons, Inc., Ney York, 2nd edition edition, 1980.
\smallskip

\noindent
R.~Twarock.
\newblock A tiling approach to virus capsid assembly explaining a structural
  puzzle in virology.
\newblock {\em J. Theor. Biol.}, 226(4):477, 2004.
\smallskip

\noindent
R.~Twarock.
\newblock Mathematical models for tubular structures in the family of
  Papovaviridae.
\newblock {\em Bull. Math. Biol.}, 67(5):973, 2005.
\smallskip

\noindent
R.~Twarock.
\newblock The architecture of viral capsids based on tiling theory.
\newblock {\em J. Theor. Med.}, 6(2):87, 2005.
\smallskip

\noindent
Zlotnick A.
\newblock To build a virus capsid: An equilibrium model of the self assembly of
  polyhedral protein complexes.
\newblock {\em J. Mol. Biol.}, 241(1):59, August 1994.

\section*{Appendix}

\subsection*{A1: The assembly tree in Zlotnick's model} 

The assembly tree consists of a single pathway, and the corresponding results are given in Table~\ref{table:ass1} adapted from (Zlotnick, 1994).
\begin{table}
  \centering
\begin{tabular}{|l||c|c|c|c|c|c|c|}
    \hline
    $n$ & Model & $O_{sym}(n)$ & $S_{n}$ & $\gamma$ & $K'_{n}$ \\ \hline
    1 & \includegraphics[width=.1\textwidth]{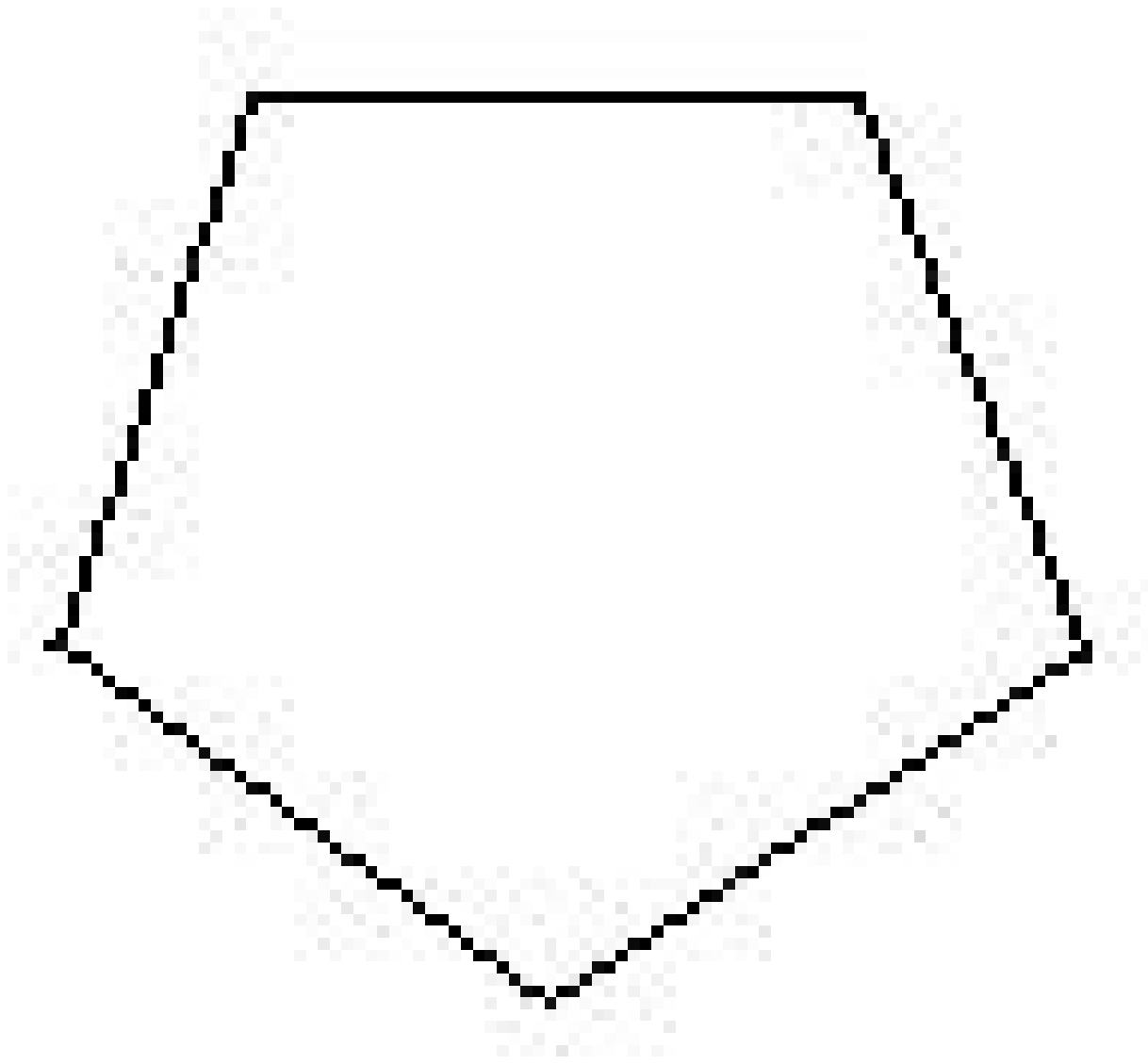} & 5 & - & - & -  \\ \hline
    2 & \includegraphics[width=.1\textwidth]{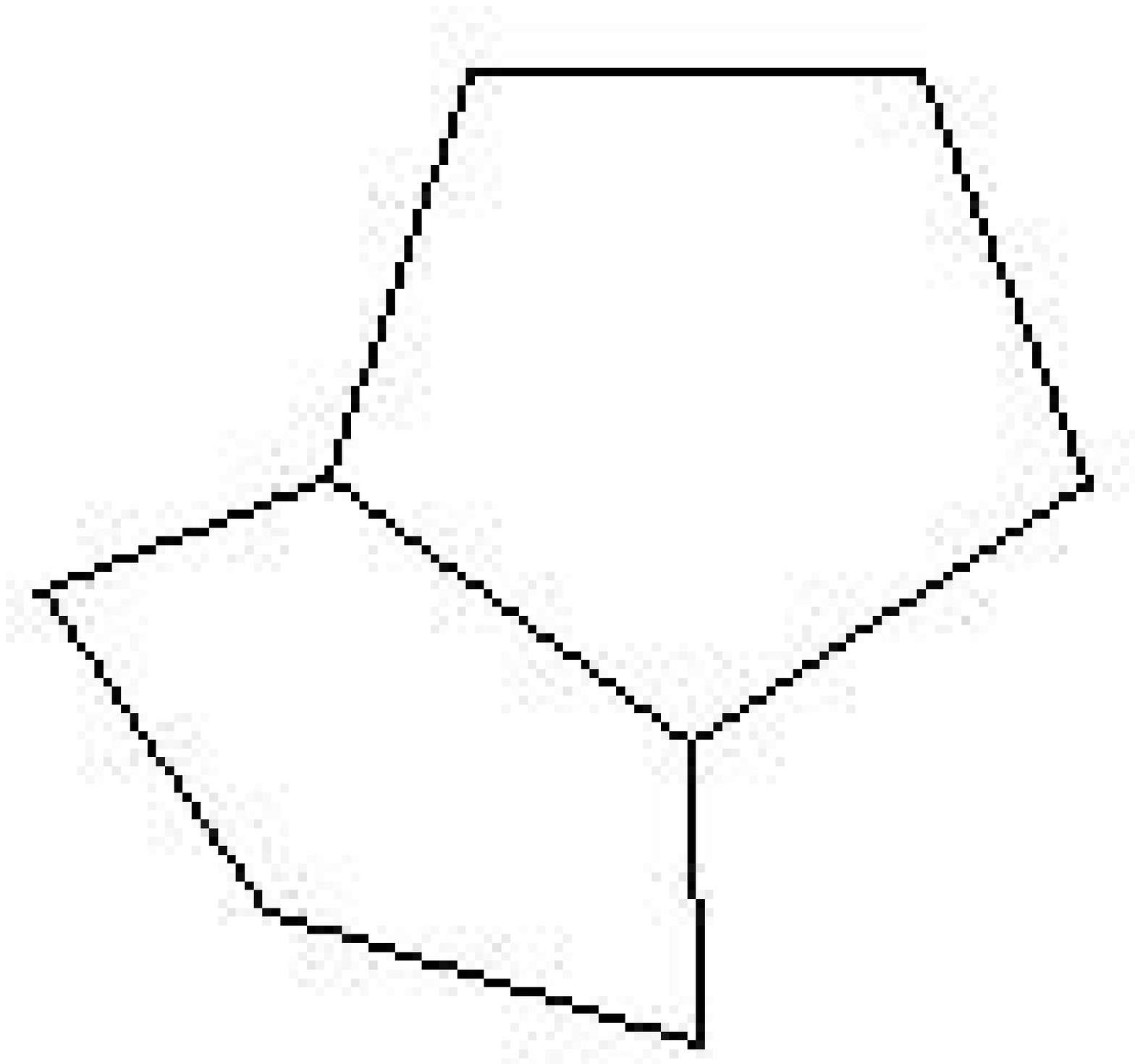} & 2 & 5/2 & 1 & $e^{-\frac{1{\Delta}G^{o}_{\gamma}}{RT}}$  \\ \hline
    3 & \includegraphics[width=.1\textwidth]{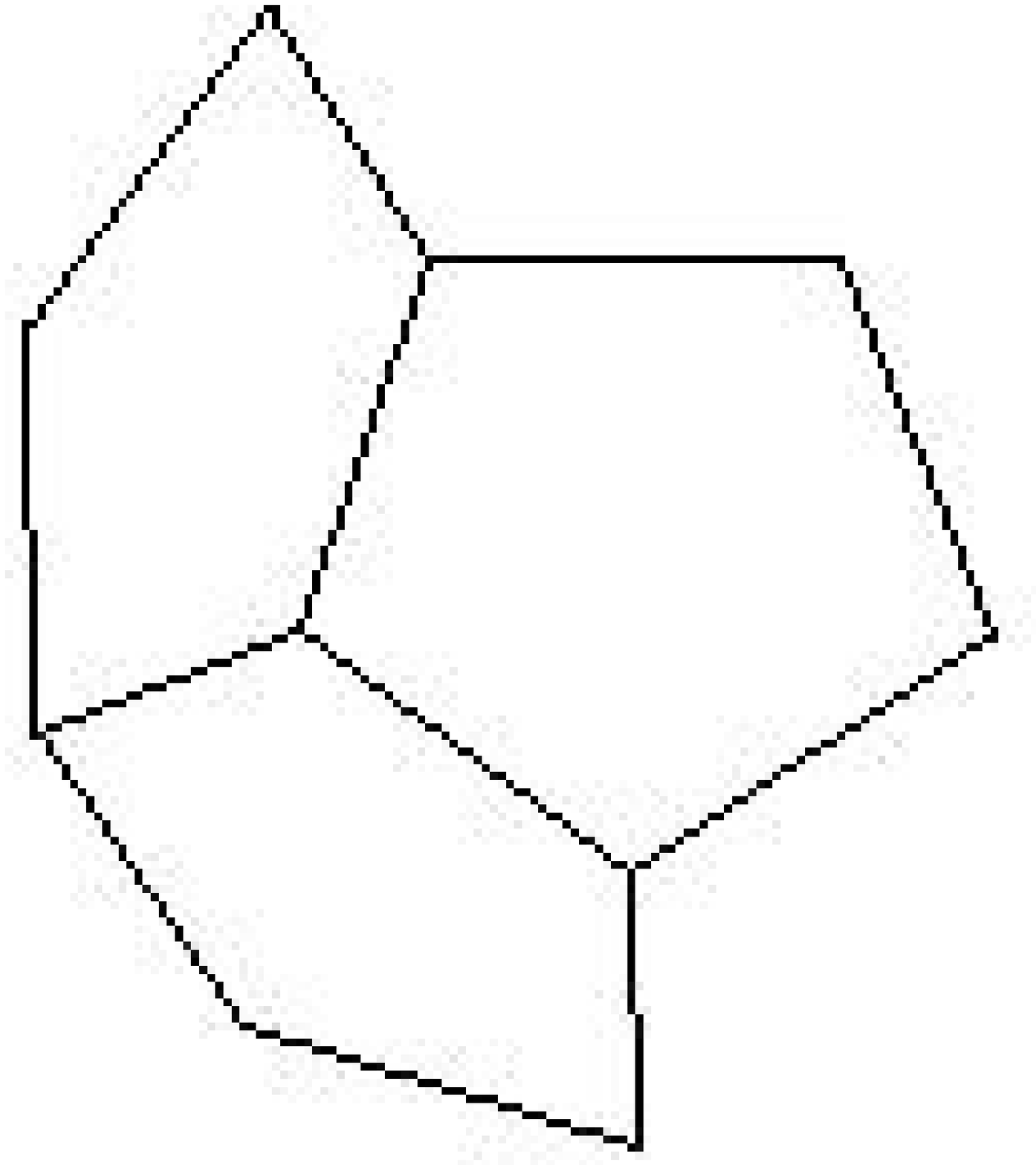} & 3 & 2/3 & 2 & $e^{-\frac{2{\Delta}G^{o}_{\gamma}}{RT}}$ \\ \hline
    4 & \includegraphics[width=.1\textwidth]{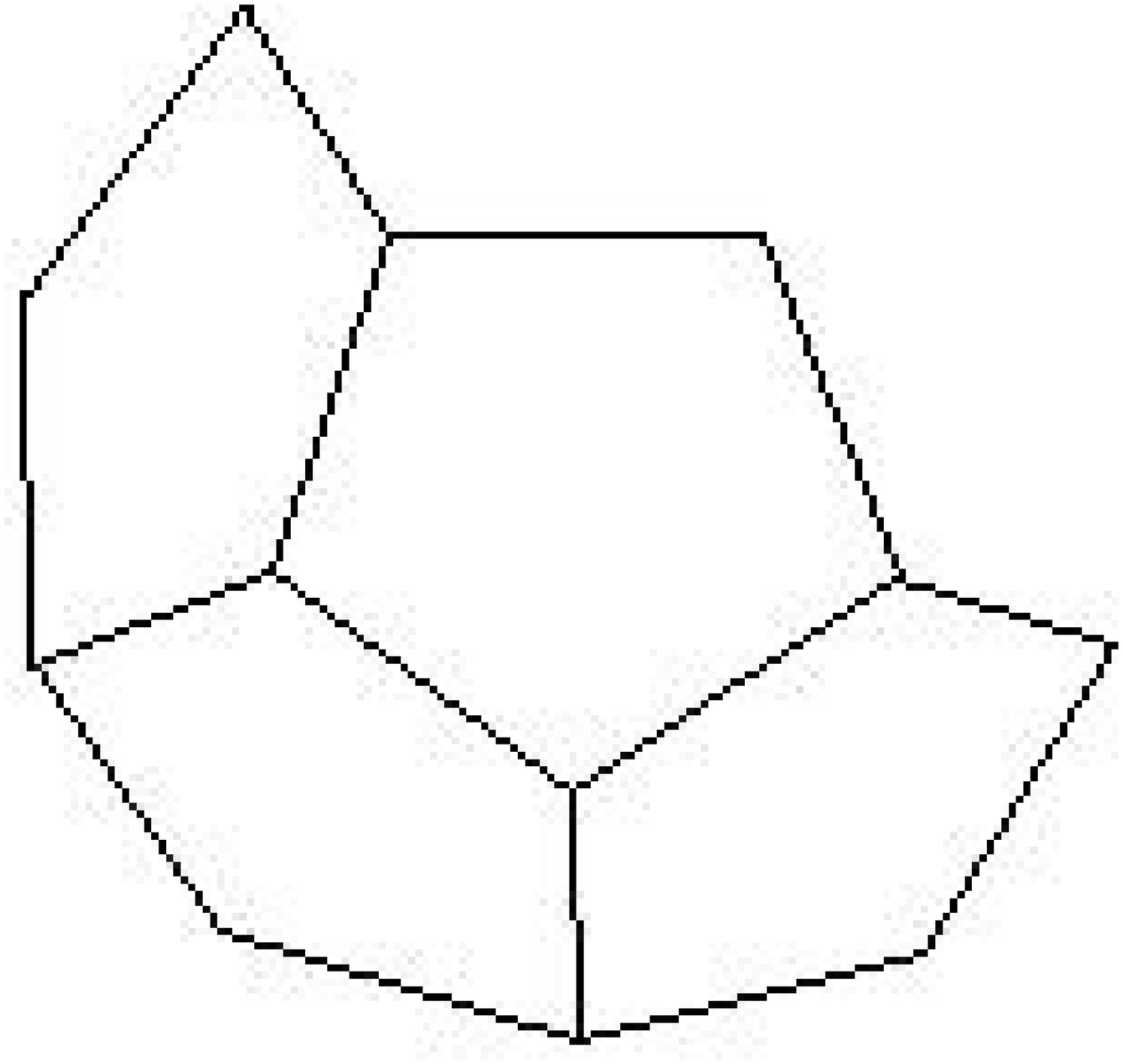} & 2 & 3/2 & 2 & $e^{-\frac{2{\Delta}G^{o}_{\gamma}}{RT}}$ \\ \hline
    5 & \includegraphics[width=.1\textwidth]{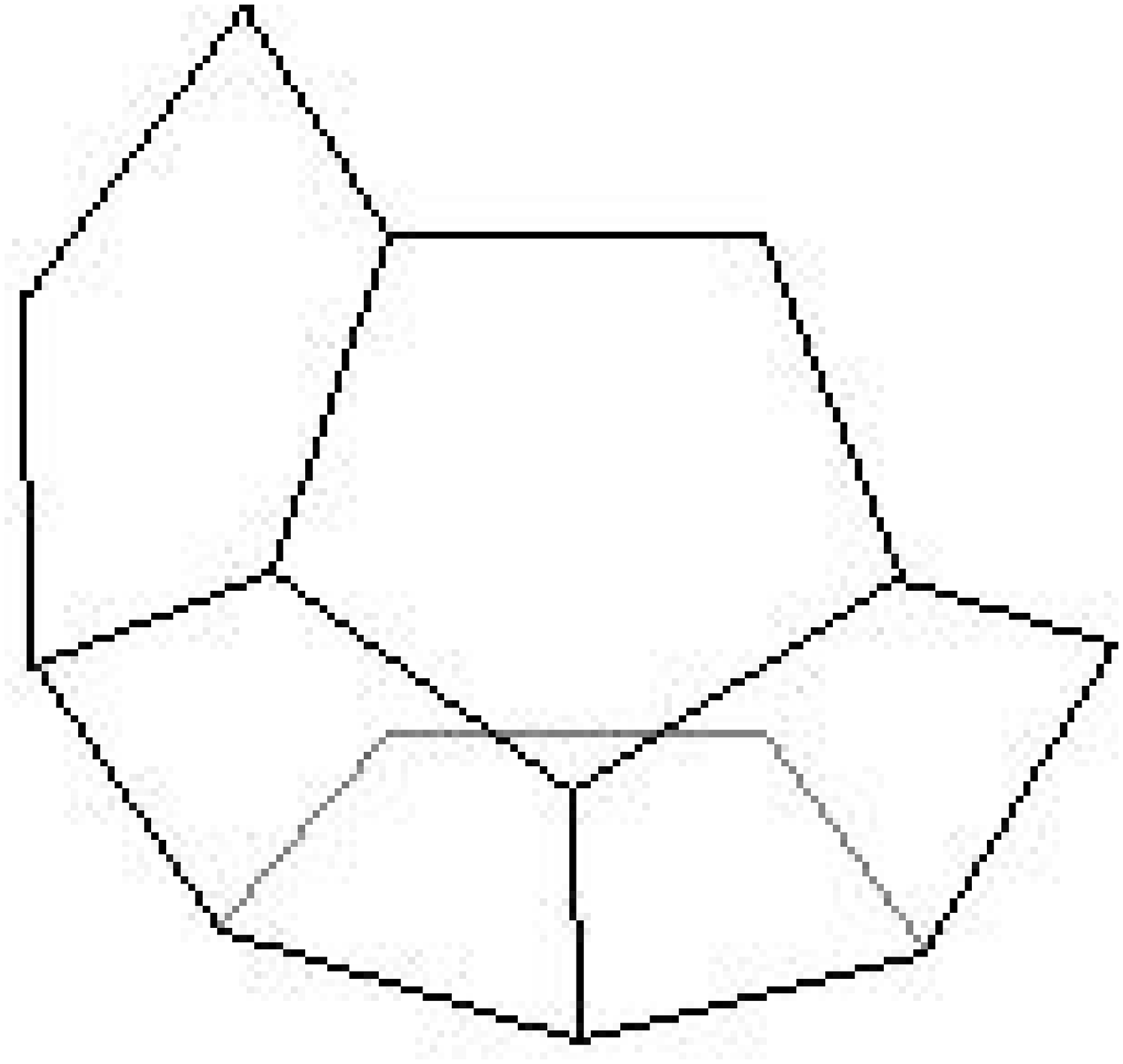} & 1 & 2 & 2 & $e^{-\frac{2{\Delta}G^{o}_{\gamma}}{RT}}$ \\ \hline
    6 & \includegraphics[width=.1\textwidth]{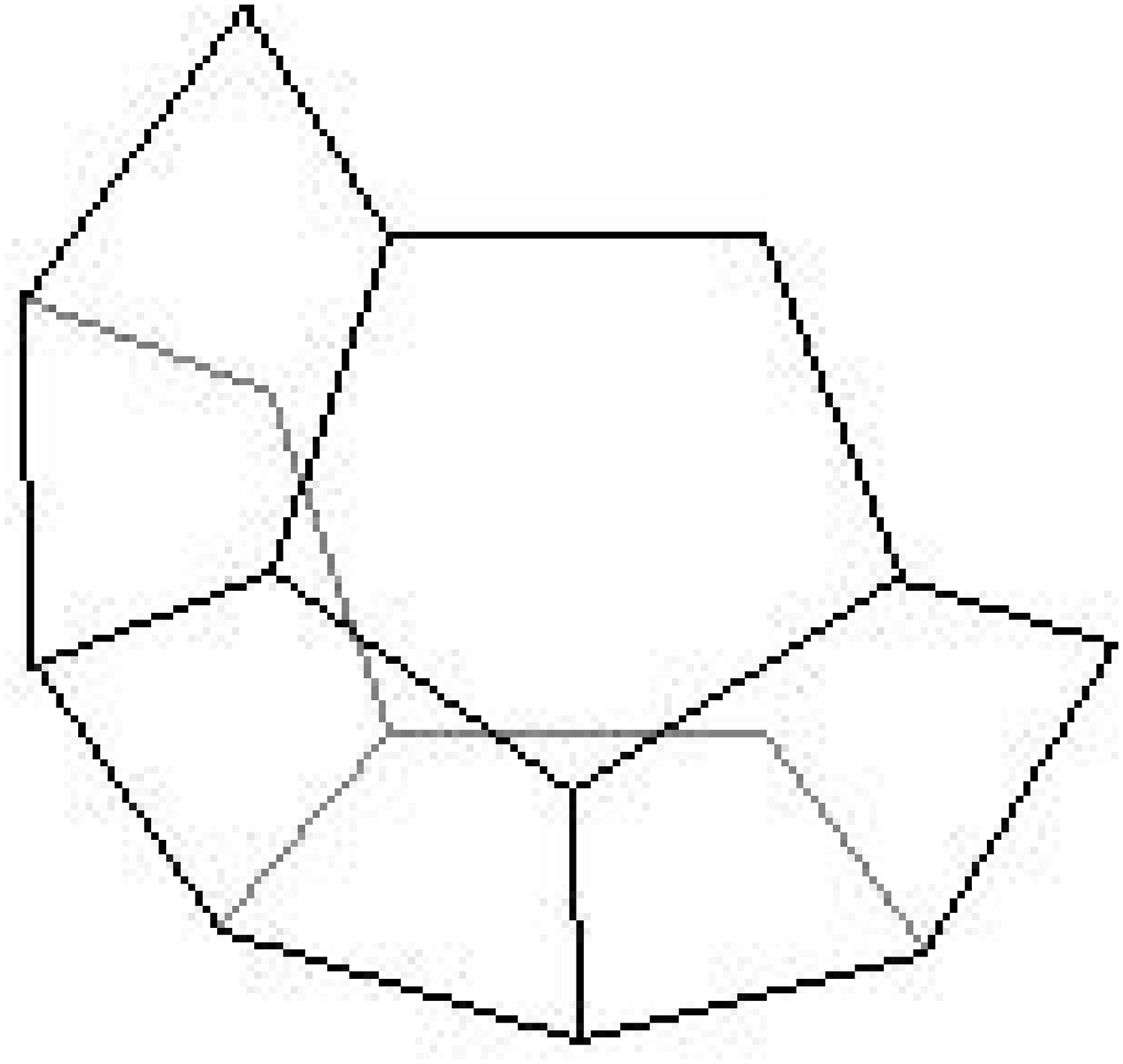} & 5 & 1/5 & 3 & $e^{-\frac{3{\Delta}G^{o}_{\gamma}}{RT}}$ \\ \hline
    7 & \includegraphics[width=.1\textwidth]{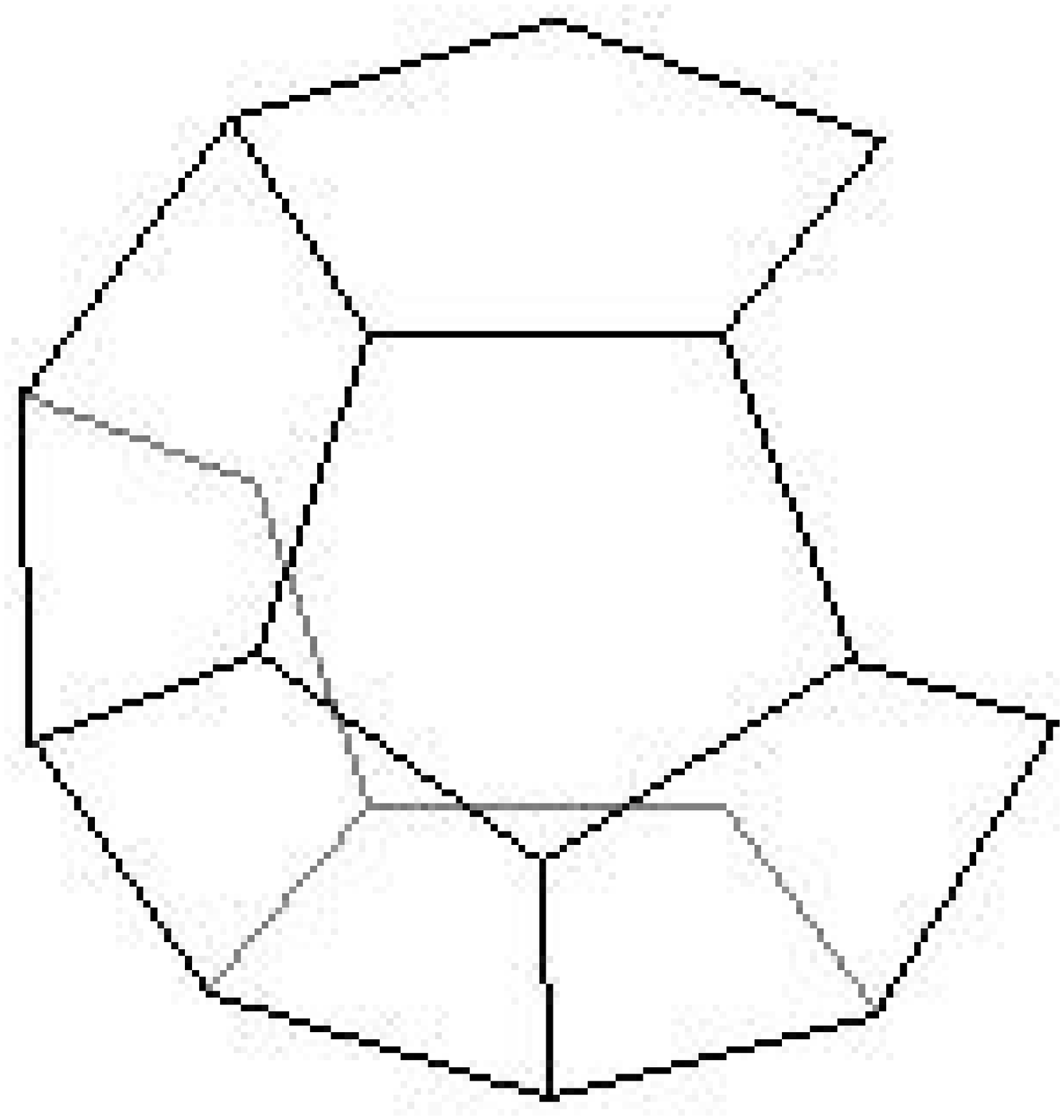} & 1 & 5 & 2 & $e^{-\frac{2{\Delta}G^{o}_{\gamma}}{RT}}$ \\ \hline
    8 & \includegraphics[width=.1\textwidth]{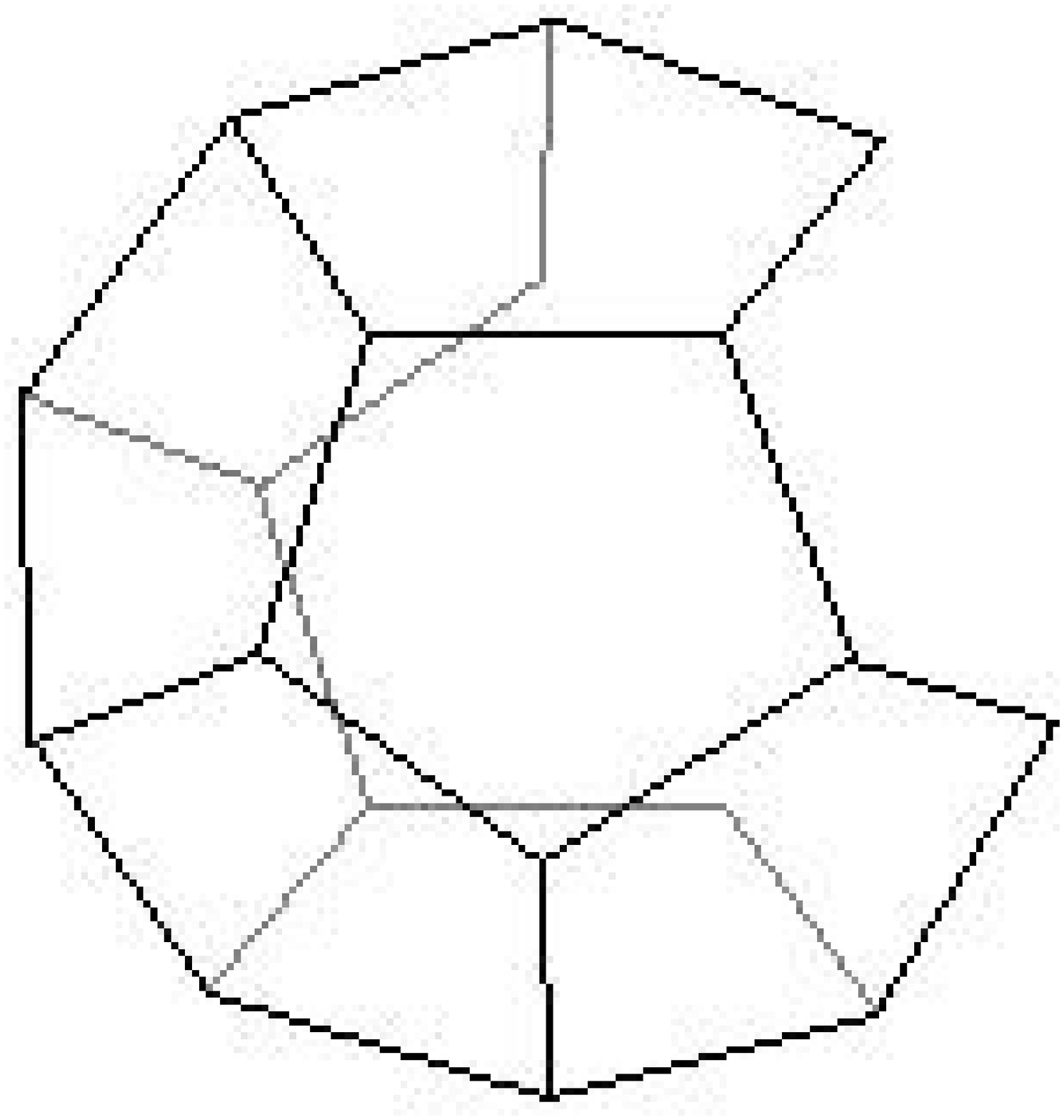} & 2 & 1/2 & 3 & $e^{-\frac{3{\Delta}G^{o}_{\gamma}}{RT}}$ \\ \hline
    9 & \includegraphics[width=.1\textwidth]{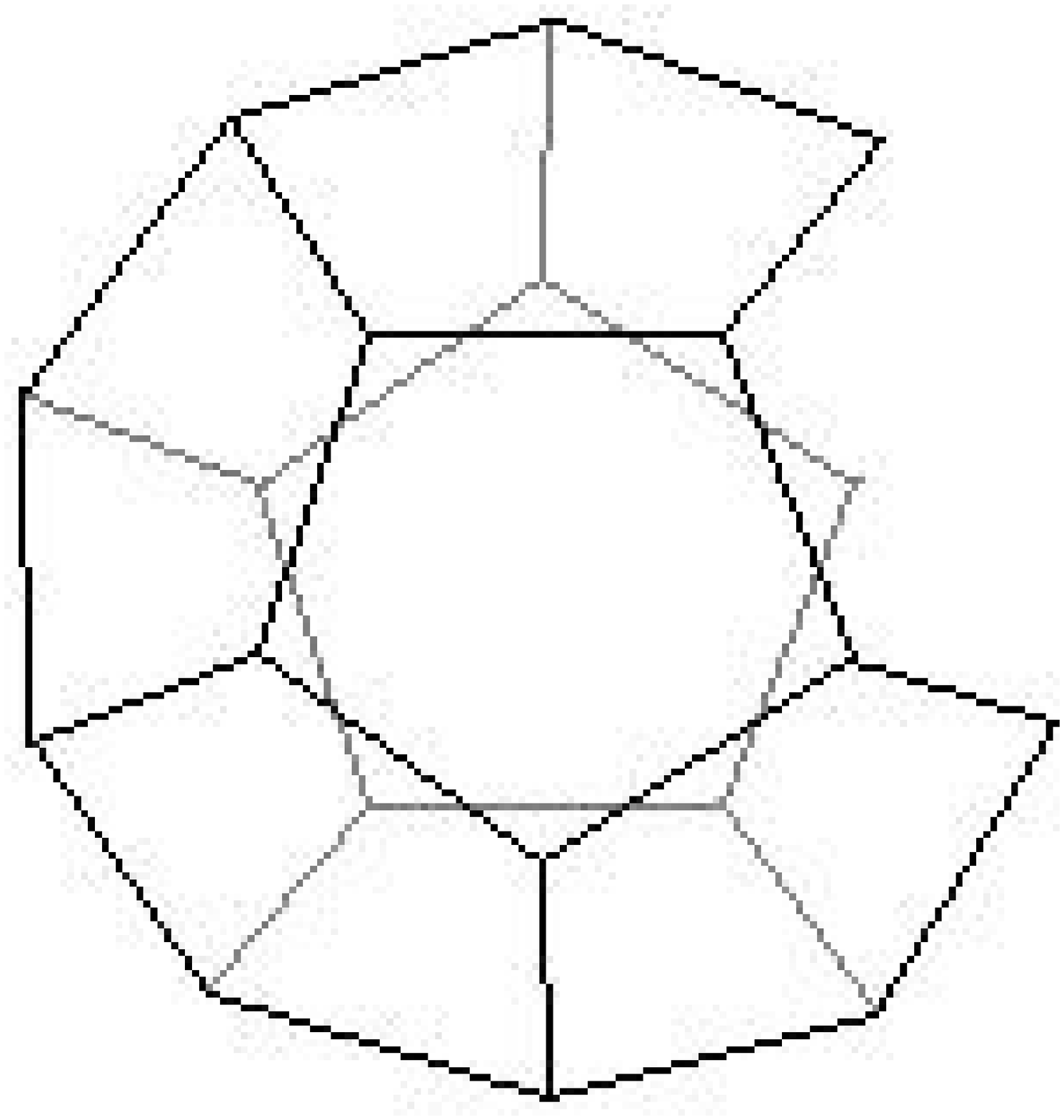} & 3 & 2/3 & 3 & $e^{-\frac{3{\Delta}G^{o}_{\gamma}}{RT}}$ \\ \hline
    10 & \includegraphics[width=.1\textwidth]{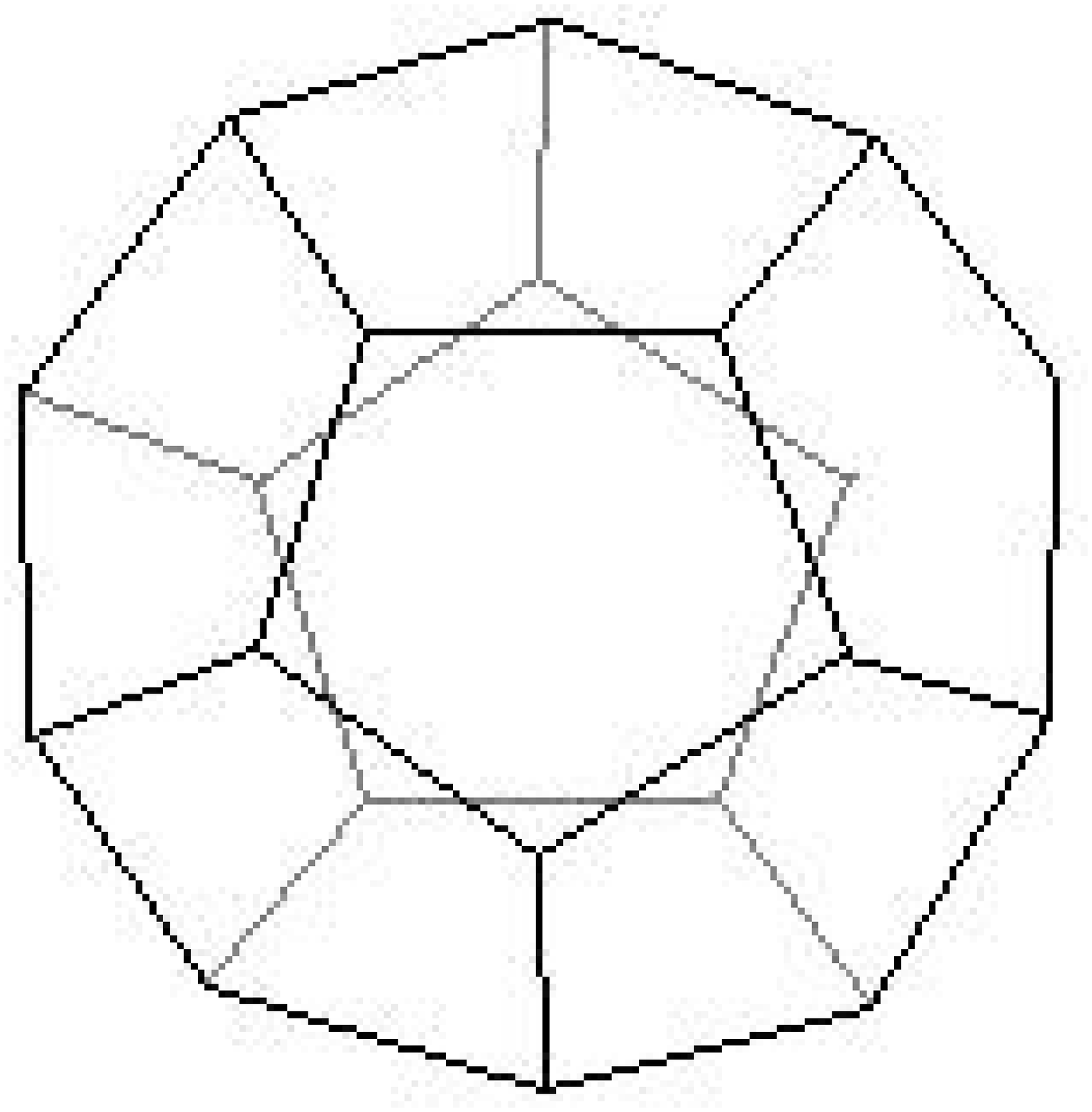} & 2 & 3/2 & 3 & $e^{-\frac{3{\Delta}G^{o}_{\gamma}}{RT}}$ \\ \hline
    11 & \includegraphics[width=.1\textwidth]{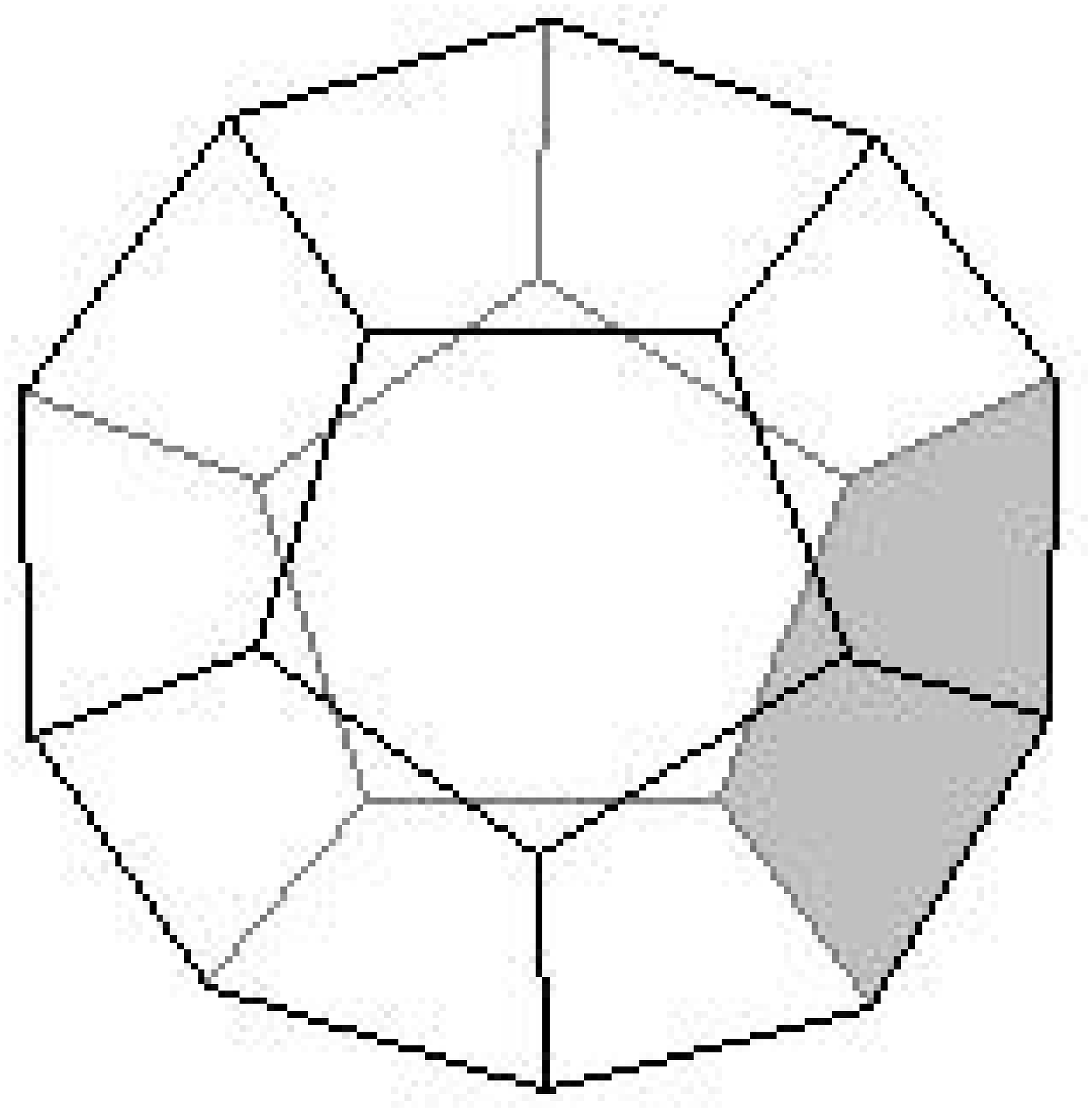} & 5 & 2/5 & 4 & $e^{-\frac{4{\Delta}G^{o}_{\gamma}}{RT}}$ \\ \hline
    12 & \includegraphics[width=.1\textwidth]{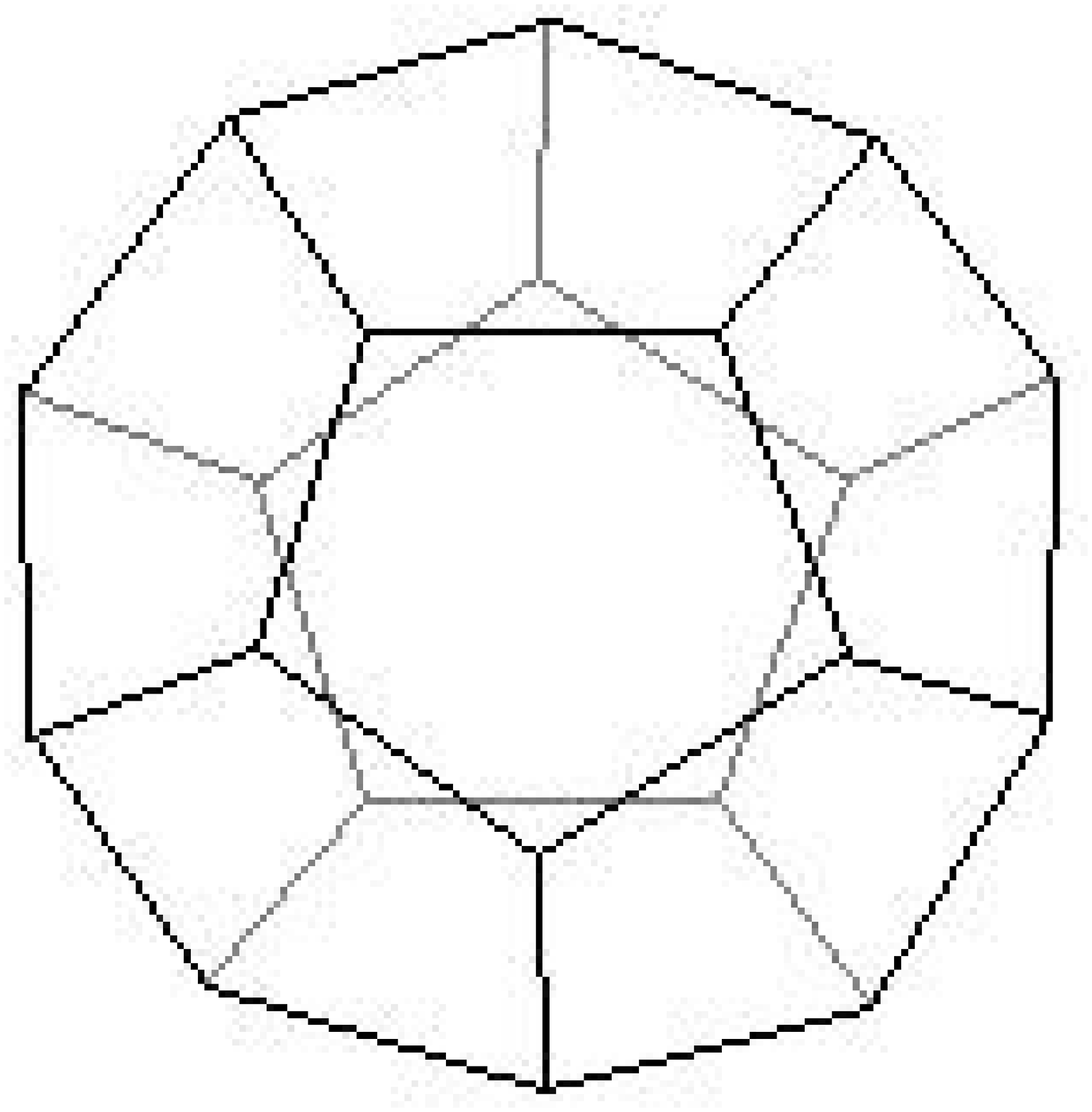} & 60 & 1/12 & 5 & $e^{-\frac{5{\Delta}G^{o}_{\gamma}}{RT}}$ \\ \hline
  \end{tabular}
  \caption{\em Assembly intermediates and factors describing their assembly, based on (Zlotnick, 1994).}\label{table:ass1}
\end{table}
\newpage 

\subsection*{A2: The assembly tree of primary intermediates in the vertex-star model} 

As discussed in Section~\ref{five}, the set of primary assembly intermediates 
depends on the association constants. 
The sequence of successive primary assembly intermediates, the building blocks added between the primary intermediates (called S and P, respectively), the number of bonds $a$, $b$ and $c$ added, as well as the order of the discrete rotational symmetries of the assembly intermediates are shown in the three tables below
for SV40.
\begin{table}
  \centering
\begin{tabular}{|l||c|c|c|c|}
    \hline
    Species & Model & New Tiles & New Bonds & $O_{sym}(n)$ \\ \hline
    1 & \includegraphics[width=.3\textwidth]{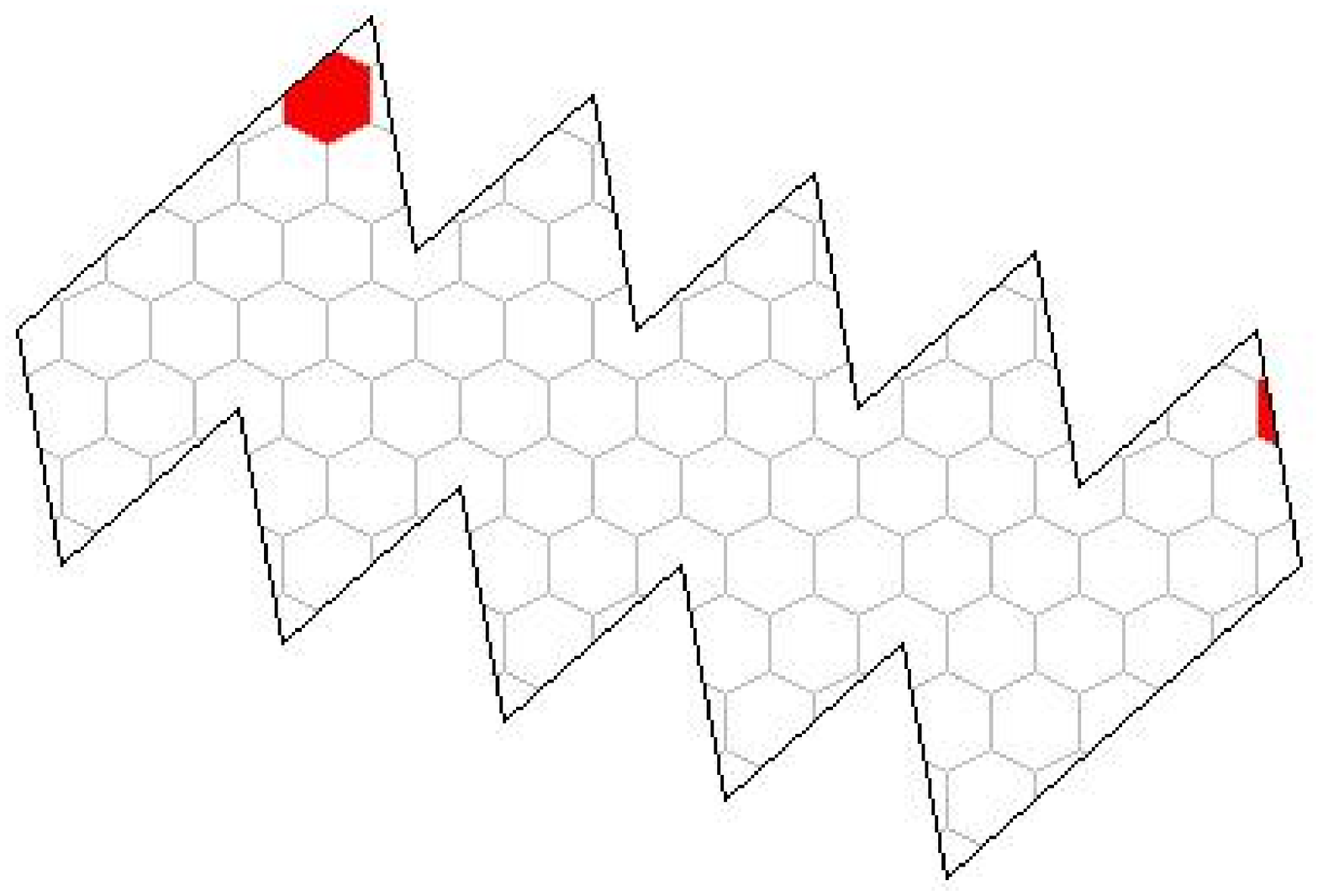} & S & - & 1  \\ \hline
    2 & \includegraphics[width=.3\textwidth]{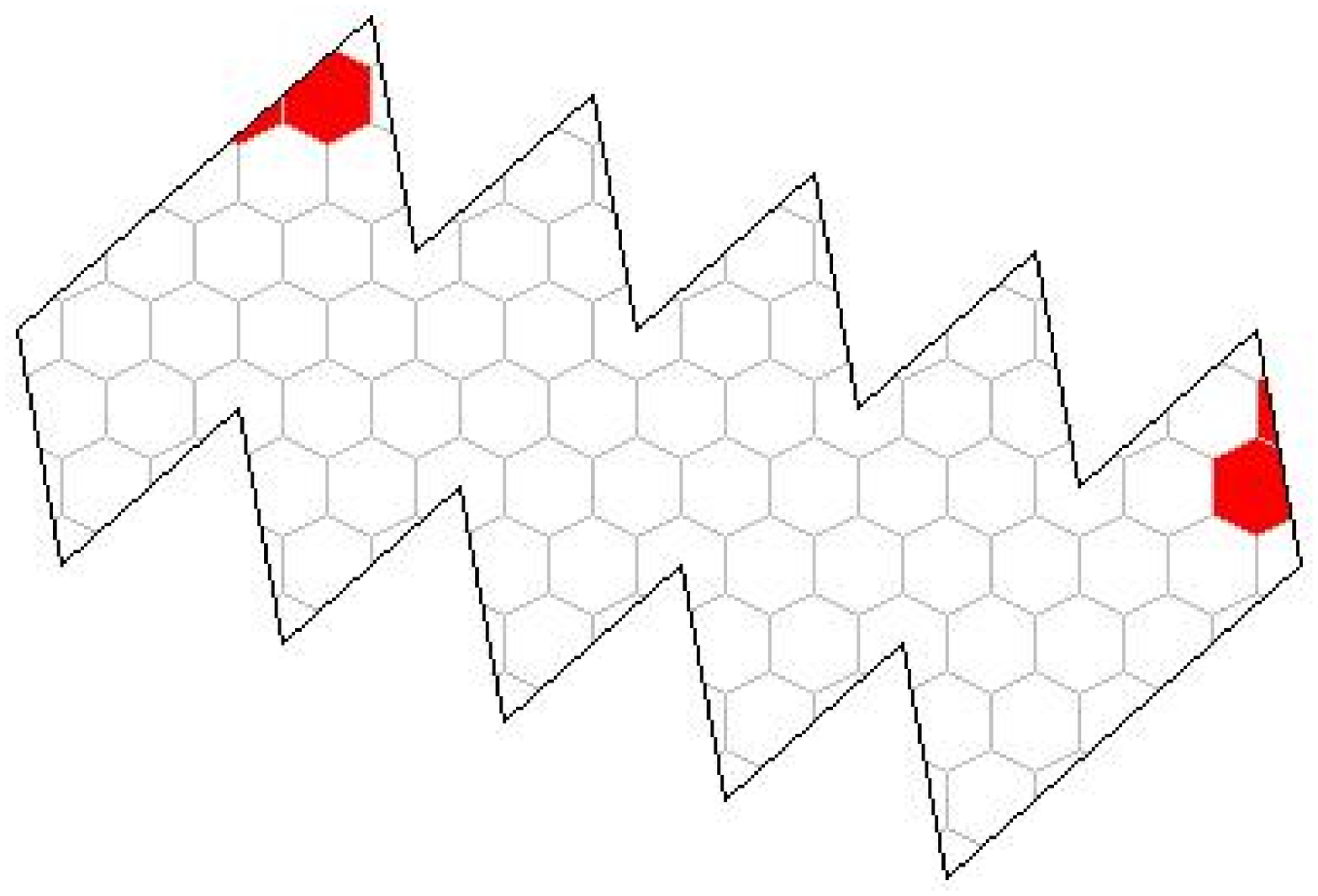} & S & $c$ & 2  \\ \hline
    3 & \includegraphics[width=.3\textwidth]{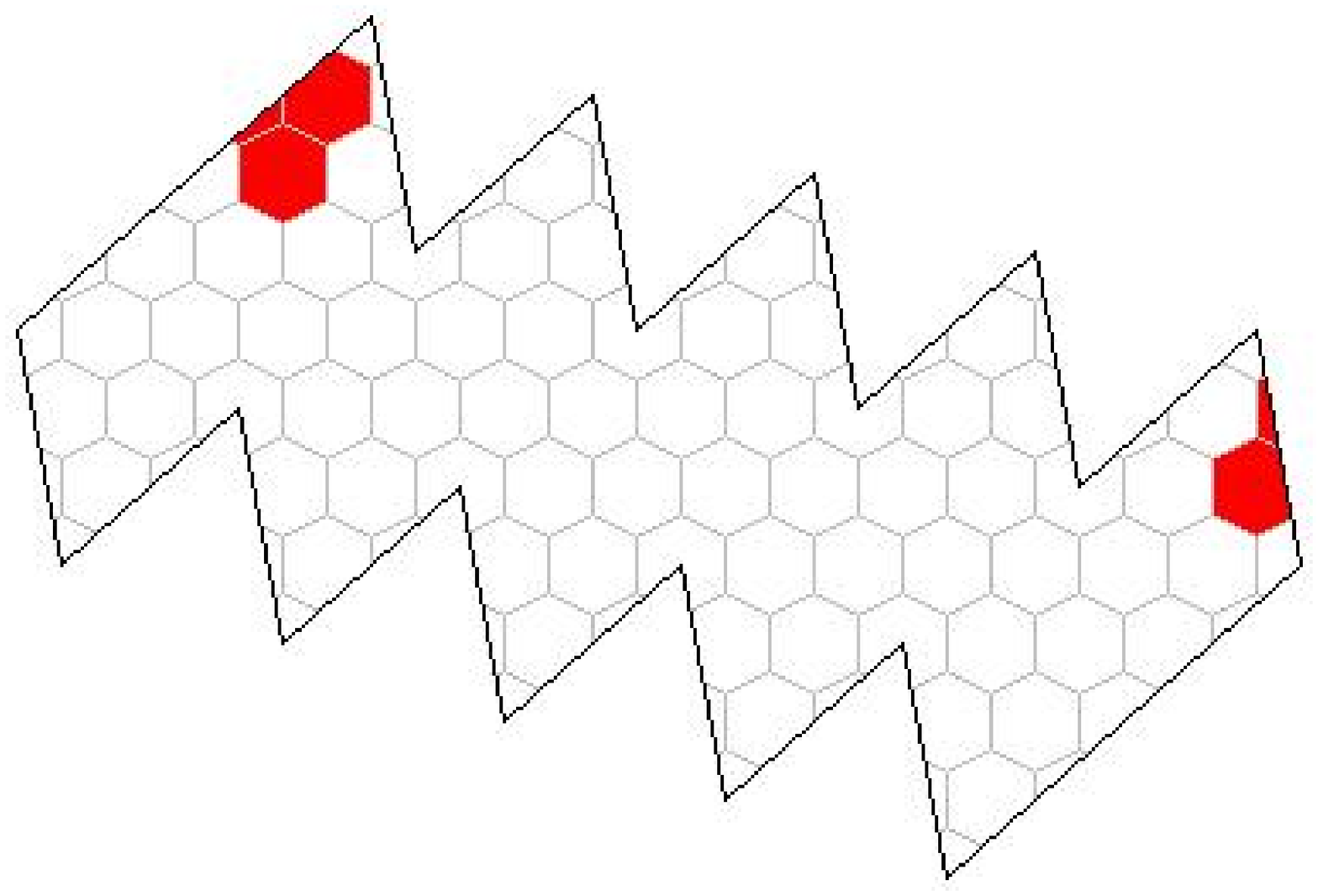} & S & $a+b$ & 1 \\ \hline
    4 & \includegraphics[width=.3\textwidth]{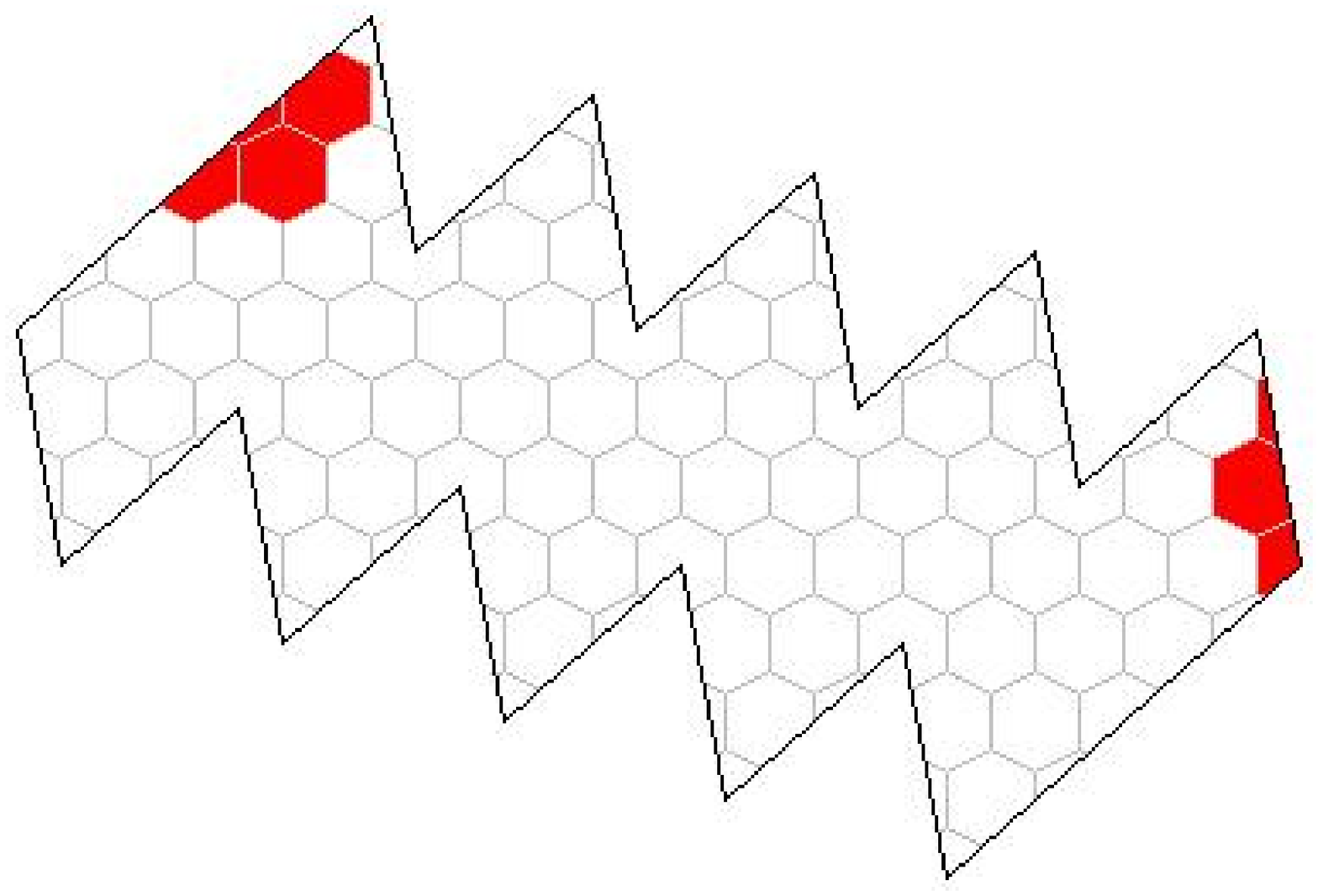} & P & $4a$ & 1 \\ \hline
    5 & \includegraphics[width=.3\textwidth]{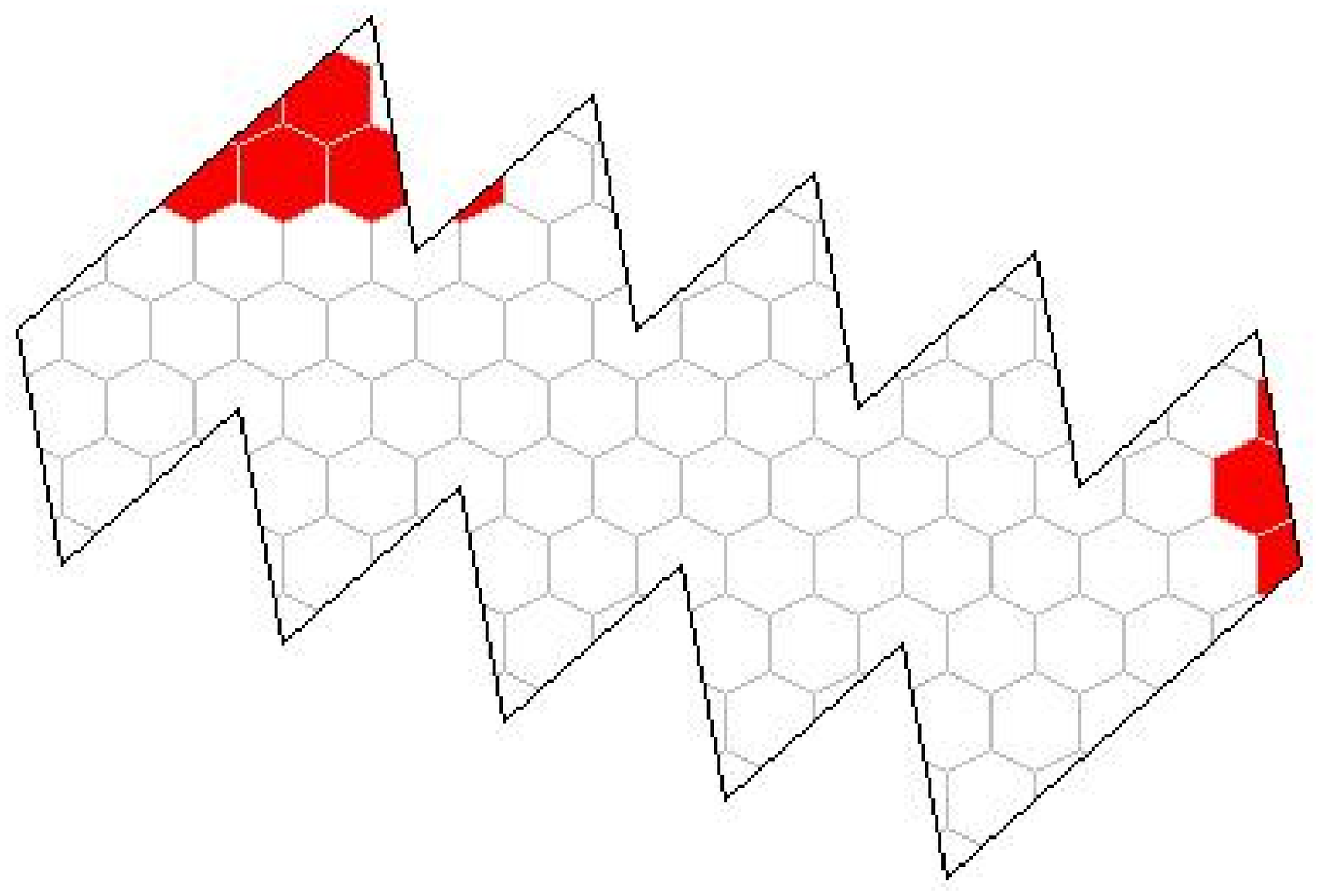} & S & $2b$ & 1 \\ \hline
    8 & \includegraphics[width=.3\textwidth]{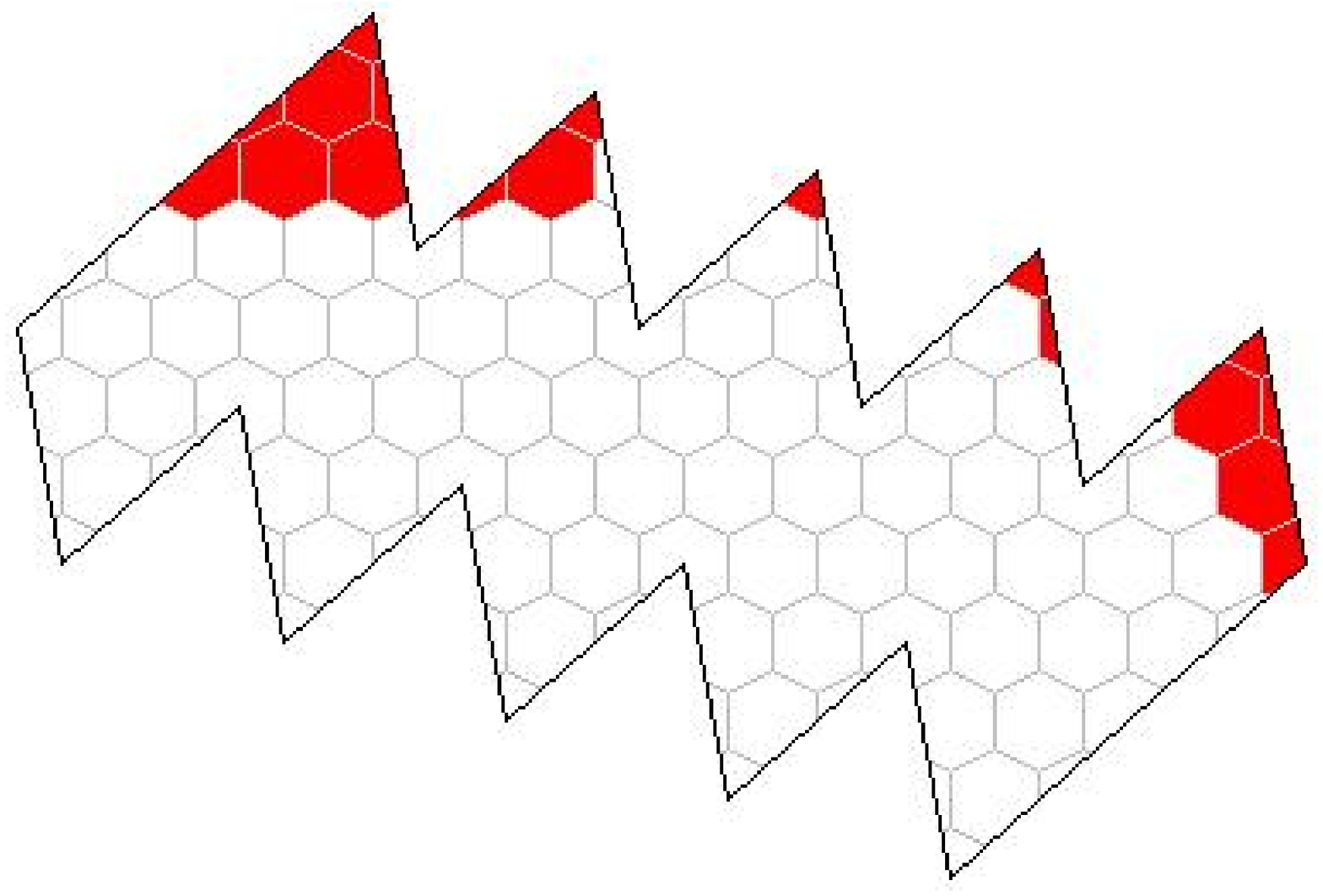} & P+2S & $8a+b+c$ & 1 \\ \hline
  \end{tabular}
  \caption{\em Assembly intermediates and factors describing their assembly.}
\label{table:polytable1}
\end{table}

\begin{table}
  \centering
\begin{tabular}{|l||c|c|c|c|}
    \hline
    Species & Model & New Tiles & New Bonds & $O_{sym}(n)$ \\ \hline
    9 & \includegraphics[width=.3\textwidth]{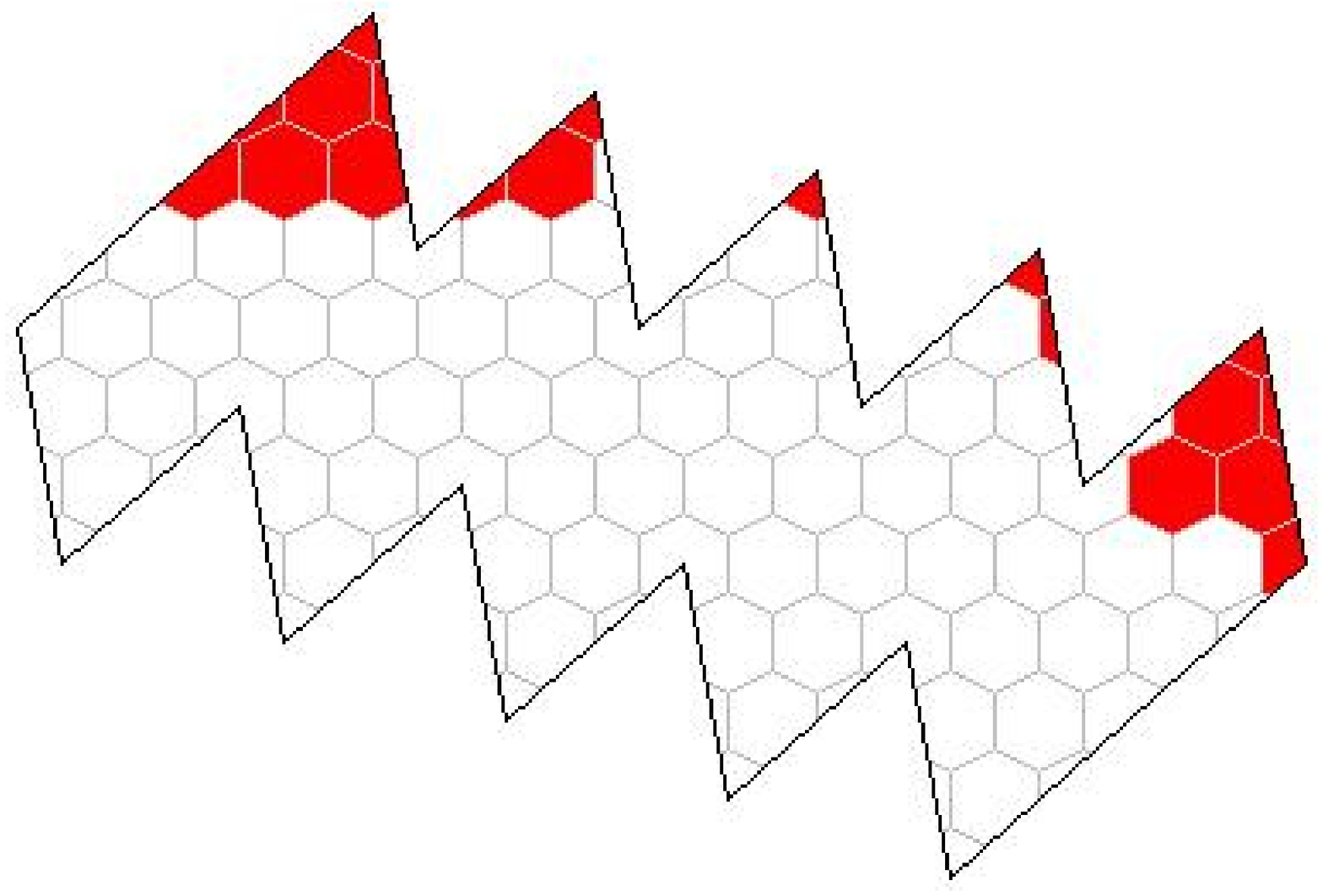} & S & $2b$ & 1  \\ \hline
    10 & \includegraphics[width=.3\textwidth]{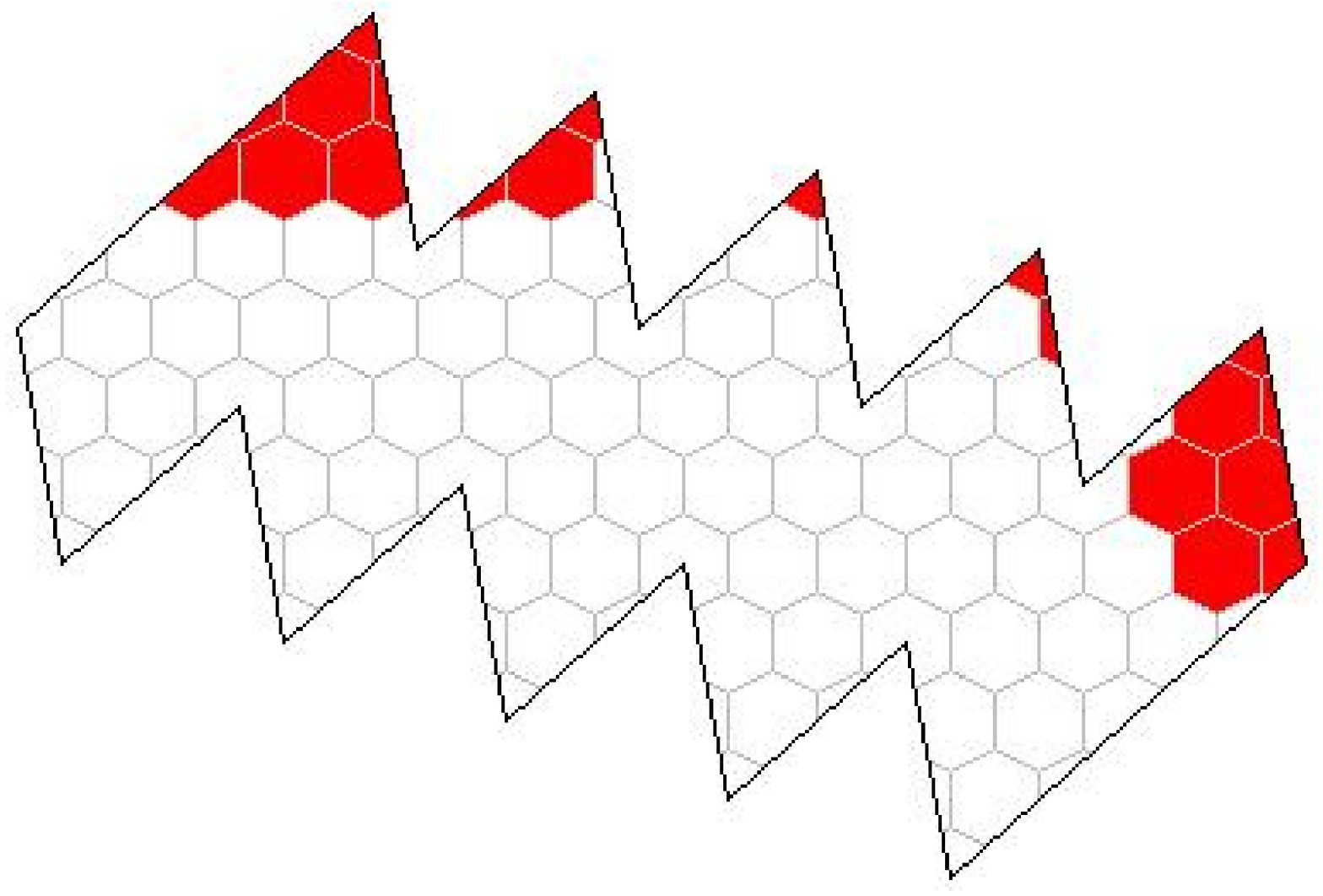} & S & $3a+c$ & 2  \\ \hline
    11 & \includegraphics[width=.3\textwidth]{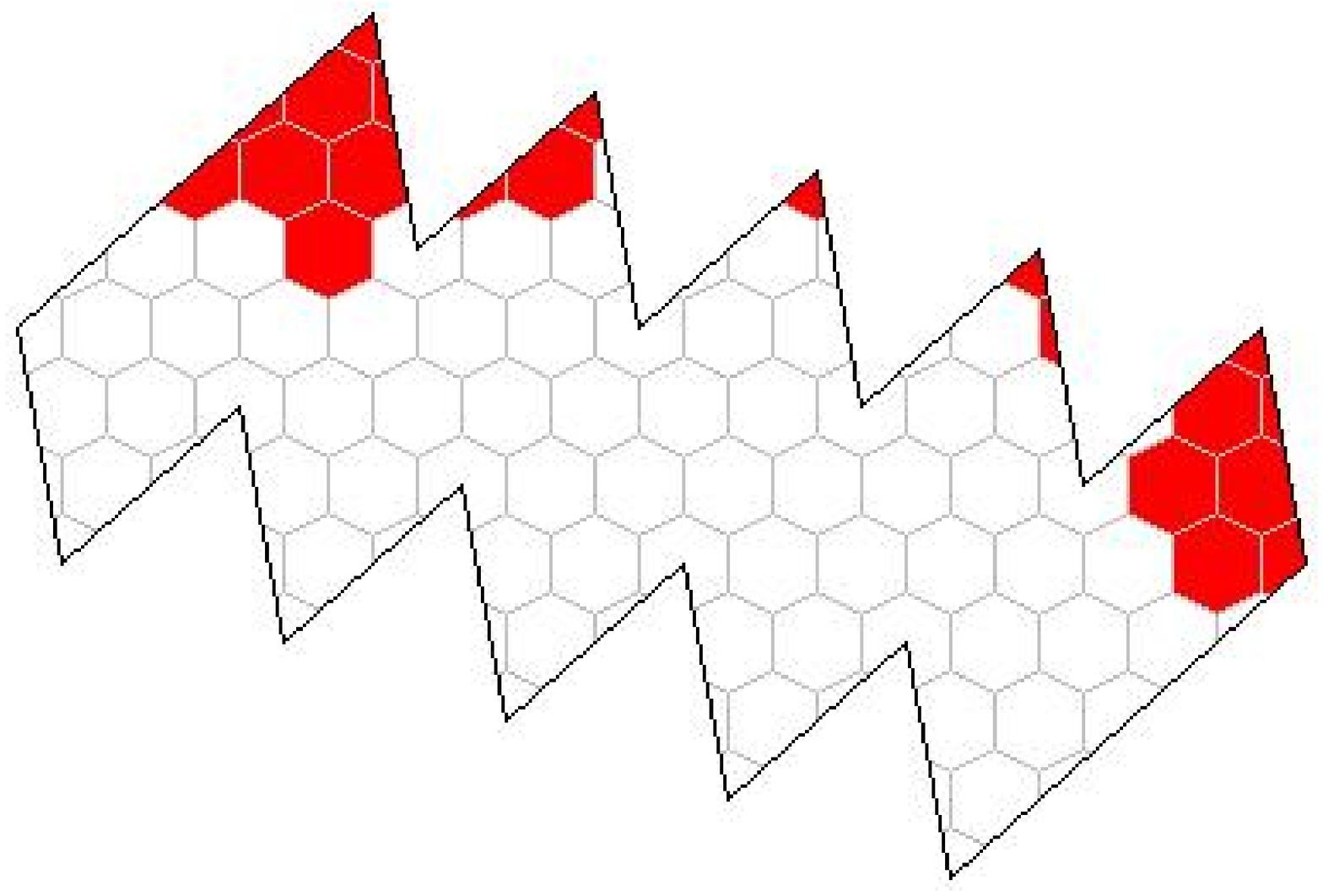} & S & $a+c$ & 1 \\ \hline
    12 & \includegraphics[width=.3\textwidth]{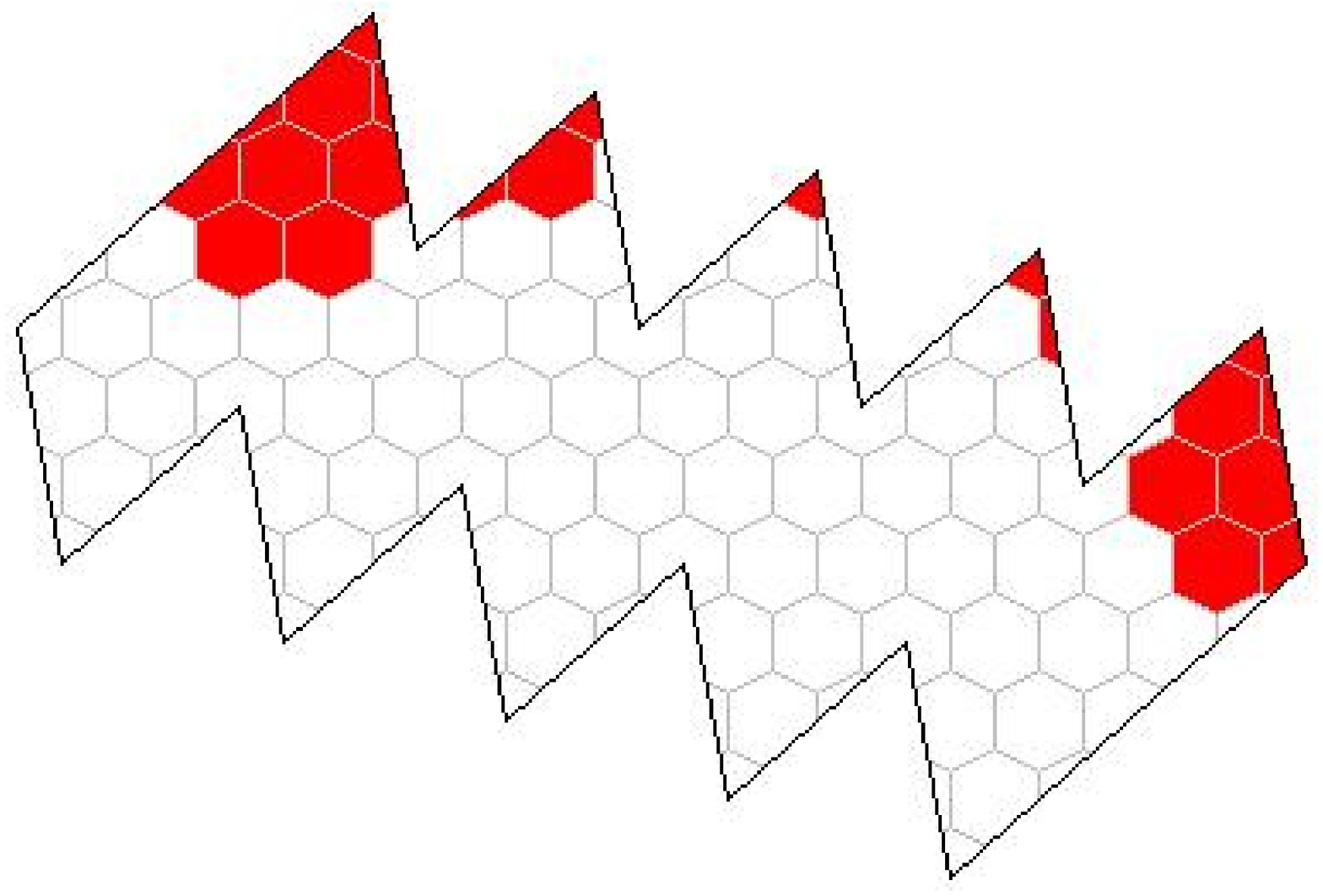} & S & $3a+b$ & 1 \\ \hline
    68 & \includegraphics[width=.3\textwidth]{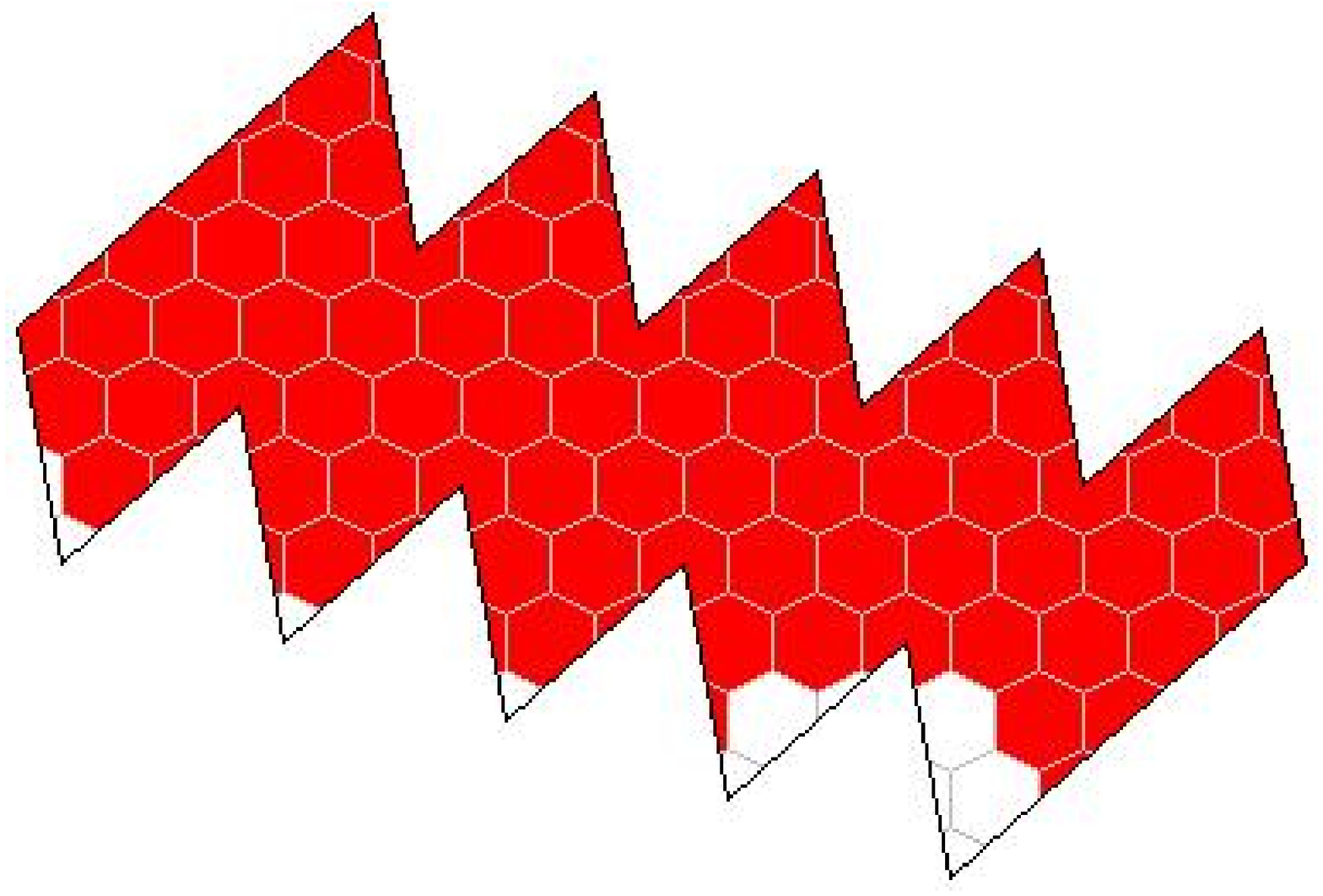} & 9P+47S & $143a+48b+24c$ & 1 \\ \hline
    69 & \includegraphics[width=.3\textwidth]{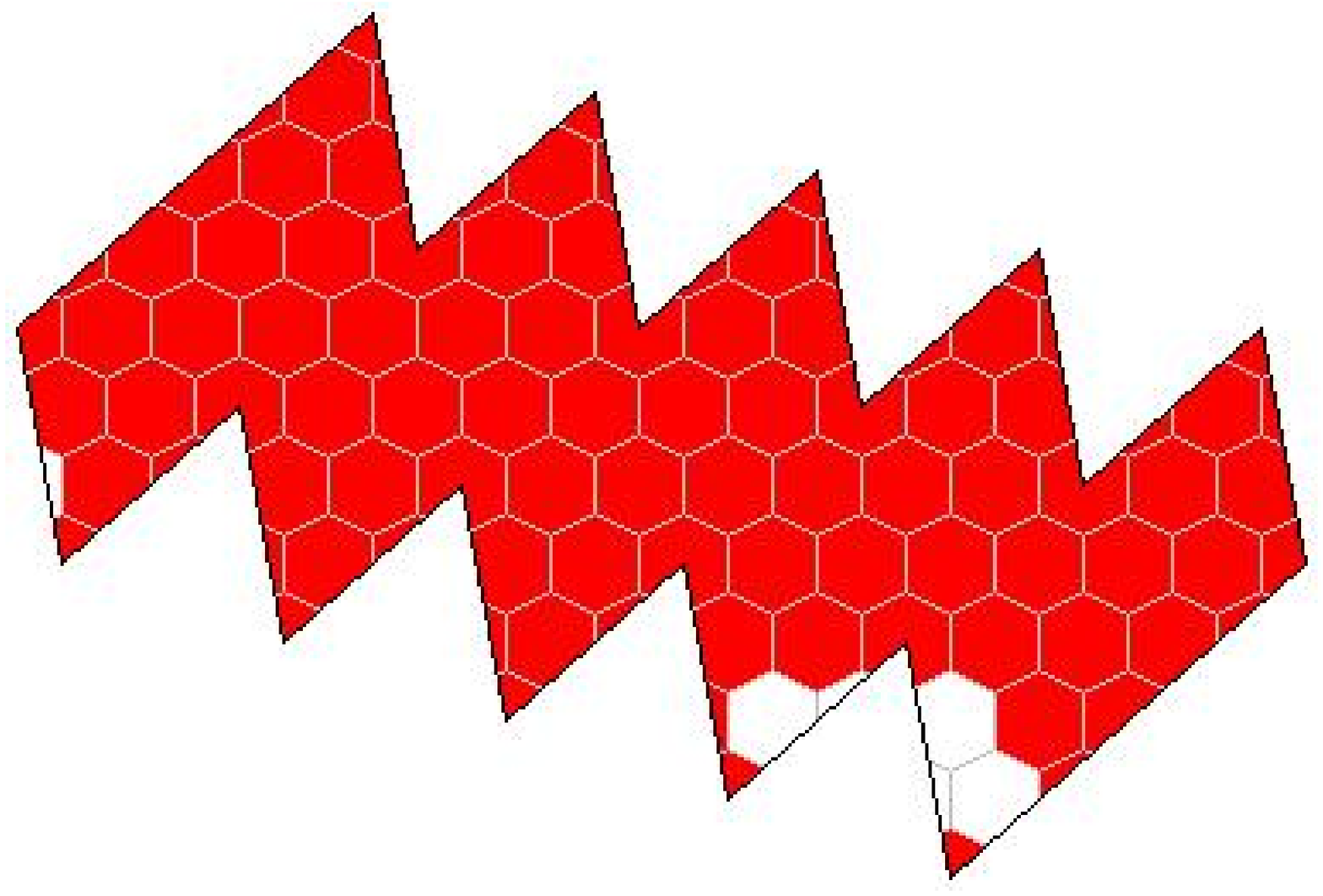} & P & $6a$ & 1 \\ \hline
  \end{tabular}
  \caption{\em Assembly intermediates and factors describing their assembly (continued).}\label{table:polytable2}
\end{table}

\begin{table}
  \centering
\begin{tabular}{|l||c|c|c|c|}
    \hline
    Species & Model & New Tiles & New Bonds & $O_{sym}(n)$ \\ \hline
    70 & \includegraphics[width=.3\textwidth]{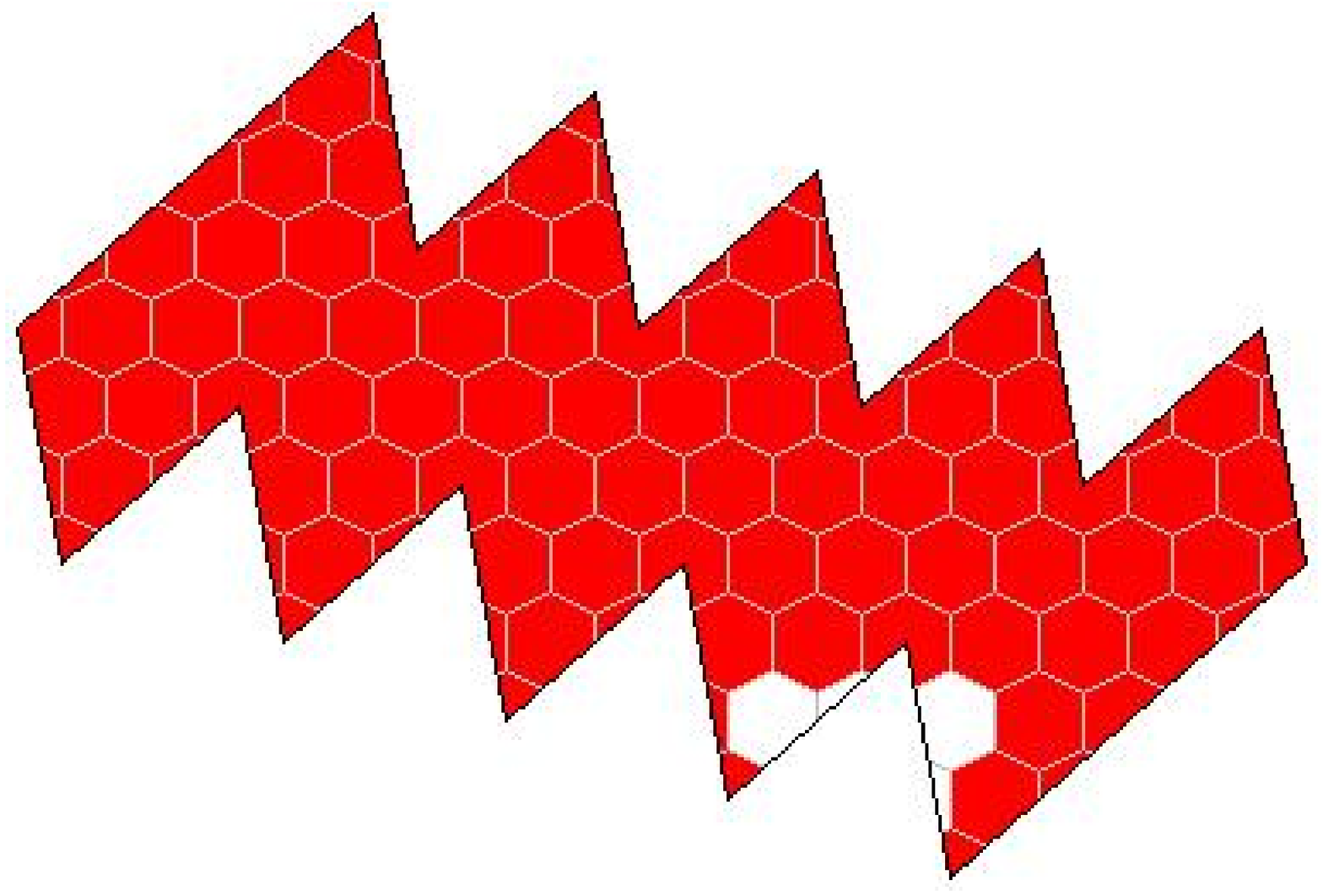} & S & $3a+b+c$ & 2  \\ \hline
    71 & \includegraphics[width=.3\textwidth]{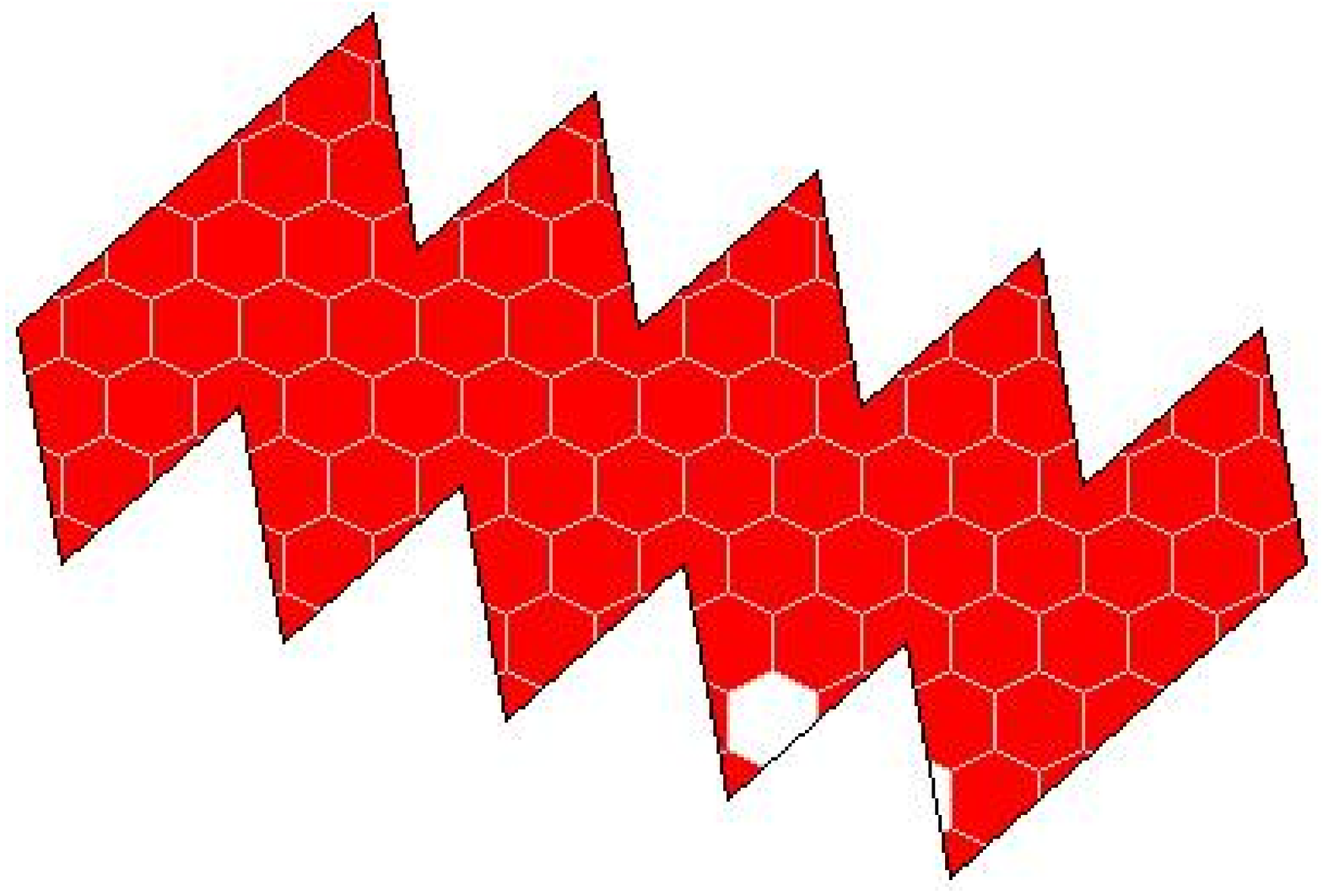} & S & $4a+2b$ & 1  \\ \hline
    72 & \includegraphics[width=.3\textwidth]{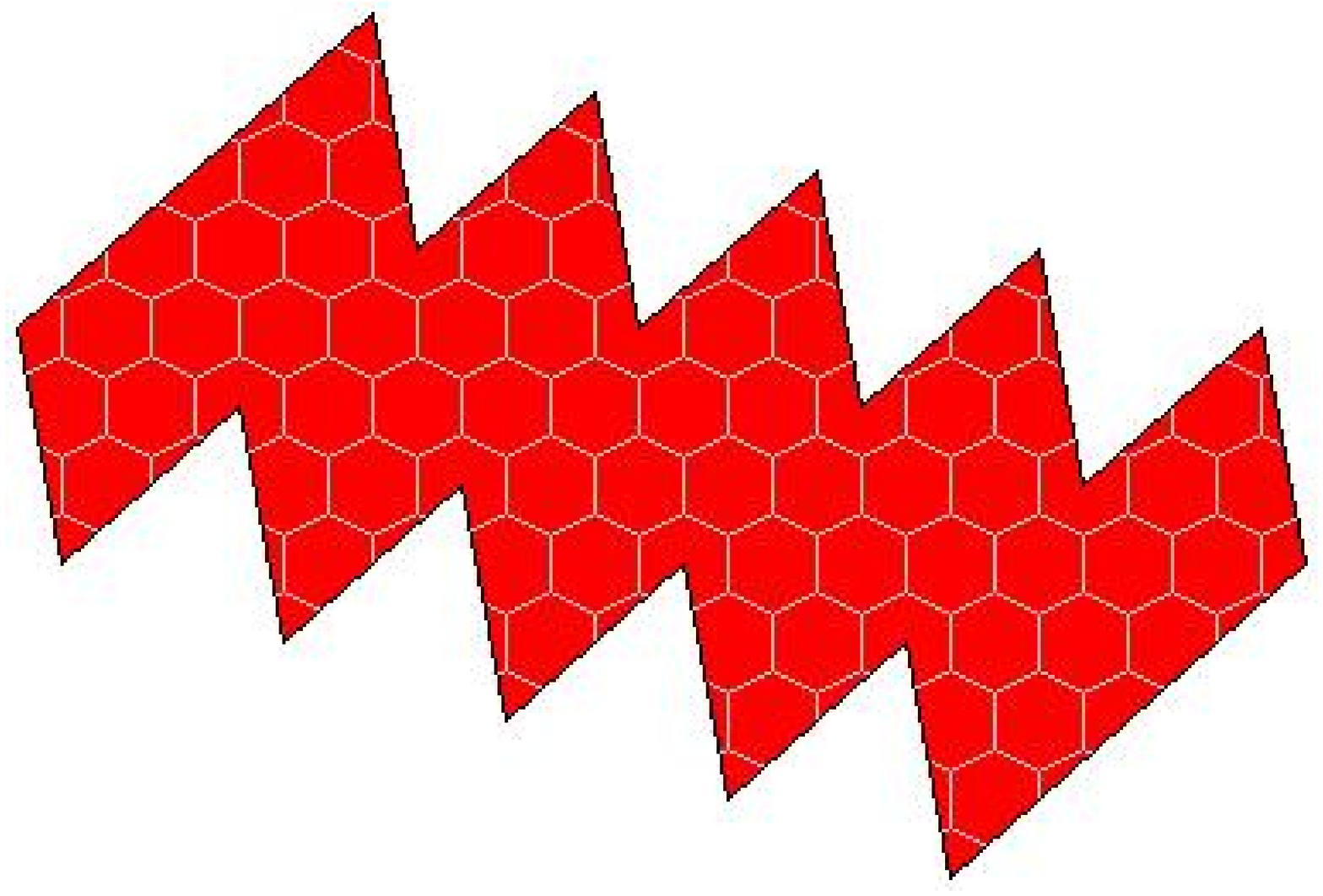} & S & $4a+2b+c$ & 60 \\ \hline
  \end{tabular}
  \caption{\em Assembly intermediates and factors describing their assembly (continued).}\label{table:polytable3}
\end{table}

\newpage 

\subsection*{A3: Primary intermediates in phase space}
In Section A3 we explore the areas in phase space in Fig.~\ref{Phase} around the shaded area corresponding to SV40. The numbers in the first row of the table correspond to the numbering of areas in that figure. For each area, a representative point is chosen and its values for $a/c$, $b/c$ and $c=1$ are given in the rows below. The column below represents in each case the occurrence of primary intermediates and their relation to primary intermediates in the other (numbered) areas of phase space. The table should be read as follows: In each row, for a given value of $N$, we enter a symbol if the corresponding intermediate is a primary intermediate, and leave the entry blank otherwise. 
For given $N$, we fill rows from left to right, starting with the heart-symbol for the first intermediate that occurs. If in another region the same intermediate occurs for this value of $N$, we enter the same symbol, otherwise, we choose another symbol from the following list (given in order of occurrence): heart, club, diamonds, spade, star.  
In this way, it can be read off which primary intermediates are shared by the areas adjacent to the red area in Fig.~\ref{Phase} representing SV40. 

\voffset -2cm

\begin{landscape}
\begin{table}
\begin{scriptsize}
\centering
\begin{tabular}{|l||c|c|c|c|c|c|c|c|c|c|c|c|c|c|c|c|c|c|c|}
    \hline
Case & SV40 & 1 & 2 & 3 & 4 & 5 & 6 & 7 & 8 & 9 & 10 &11 &12 &13 &14 &15 &16 &17 &18 \\ \hline
    $a/c$ & 0.47 & 0.45 & 0.54 & 0.6 & 0.65 & 0.65 & 0.54 & 0.48  & 0.46 & 0.4 & 0.38 & 0.35 & 0.32 & 0.3 & 0.28 & 0.28 & 0.32 & 0.36 & 0.42 \\ \hline
    $b/c$ & 0.92 & 0.75 & 0.9 & 1.08 & 1.2 & 1.4 & 1.2 &  1.1 & 0.99 & 0.85 & 0.86 & 0.8 & 0.7 & 0.62 & 0.5 & 0.46 & 0.5 & 0.65 & 0.67 \\ \hline \hline
    N & \multicolumn{19}{|c|}{Primary Node}  \\ \hline
    1 & ${\color{red}\heartsuit}$ & ${\color{red}\heartsuit}$ & ${\color{red}\heartsuit}$ & ${\color{red}\heartsuit}$ & ${\color{red}\heartsuit}$ & ${\color{red}\heartsuit}$ & ${\color{red}\heartsuit}$ &  ${\color{red}\heartsuit}$ & ${\color{red}\heartsuit}$ &${\color{red}\heartsuit}$ &
${\color{red}\heartsuit}$ & ${\color{red}\heartsuit}$ & ${\color{red}\heartsuit}$ & ${\color{red}\heartsuit}$ & ${\color{red}\heartsuit}$ & ${\color{red}\heartsuit}$ & ${\color{red}\heartsuit}$ & ${\color{red}\heartsuit}$ & ${\color{red}\heartsuit}$  \\ \hline
    2 & ${\color{red}\heartsuit}$ & ${\color{red}\heartsuit}$ & $\clubsuit$ & $\clubsuit$ & $\clubsuit$ & ${\color{red}\diamondsuit}$ & ${\color{red}\diamondsuit}$ &   ${\color{red}\diamondsuit}$ & ${\color{red}\heartsuit}$ &${\color{red}\heartsuit}$ &
${\color{red}\heartsuit}$ & ${\color{red}\heartsuit}$ & ${\color{red}\heartsuit}$ & ${\color{red}\heartsuit}$ & ${\color{red}\heartsuit}$ & ${\color{red}\heartsuit}$ & ${\color{red}\heartsuit}$ & ${\color{red}\heartsuit}$ & ${\color{red}\heartsuit}$   \\ \hline
    3 & ${\color{red}\heartsuit}$ & ${\color{red}\heartsuit}$ &  $\clubsuit$ &  $\clubsuit$ & $\clubsuit$ & ${\color{red}\diamondsuit}$ & ${\color{red}\diamondsuit}$ &  ${\color{red}\diamondsuit}$ &${\color{red}\heartsuit}$ & ${\color{red}\heartsuit}$ &
${\color{red}\heartsuit}$ & ${\color{red}\heartsuit}$ & ${\color{red}\heartsuit}$ & ${\color{red}\heartsuit}$ & ${\color{red}\heartsuit}$ & ${\color{red}\heartsuit}$ & ${\color{red}\heartsuit}$ & ${\color{red}\heartsuit}$ & ${\color{red}\heartsuit}$\\ \hline
    4 & ${\color{red}\heartsuit}$ & ${\color{red}\heartsuit}$ &  ${\color{red}\heartsuit}$ &    ${\color{red}\heartsuit}$ & ${\color{red}\heartsuit}$ & $\clubsuit$ & $\clubsuit$ &  $\clubsuit$ &$\clubsuit$ & $\clubsuit$ &
$\clubsuit$ & $\clubsuit$& $\clubsuit$ & $\clubsuit$ & ${\color{red}\heartsuit}$ & ${\color{red}\heartsuit}$ & ${\color{red}\heartsuit}$ & ${\color{red}\heartsuit}$ & ${\color{red}\heartsuit}$ \\ \hline
    5 & ${\color{red}\heartsuit}$ & ${\color{red}\heartsuit}$ & ${\color{red}\heartsuit}$ &     ${\color{red}\heartsuit}$ & ${\color{red}\heartsuit}$ & ${\color{red}\heartsuit}$ & ${\color{red}\heartsuit}$ &  ${\color{red}\heartsuit}$ &  ${\color{red}\heartsuit}$ &${\color{red}\heartsuit}$&
${\color{red}\heartsuit}$ & ${\color{red}\heartsuit}$ & $\clubsuit$ & $\clubsuit$ &  & ${\color{red}\diamondsuit}$ &  & ${\color{red}\heartsuit}$ & ${\color{red}\heartsuit}$  \\ \hline
    6 &  &  &    &       &  & ${\color{red}\heartsuit}$ & ${\color{red}\heartsuit}$ &  ${\color{red}\heartsuit}$ &   &  
&  & & $\clubsuit$ & $\clubsuit$ &  & ${\color{red}\diamondsuit}$ &  &  &
\\ \hline
    7 &  &  &    &      &  & ${\color{red}\heartsuit}$ & ${\color{red}\heartsuit}$ &   ${\color{red}\heartsuit}$&  &   
&  & & $\clubsuit$ & $\clubsuit$ &  & ${\color{red}\diamondsuit}$ &  &  & 
\\ \hline
    8 & ${\color{red}\heartsuit}$ & ${\color{red}\heartsuit}$ &  &       &  & $\clubsuit$ & $\clubsuit$ &   $\clubsuit$ & & 
 &  &  & ${\color{red}\diamondsuit}$ & ${\color{red}\diamondsuit}$ & ${\color{red}\diamondsuit}$ & ${\color{red}\diamondsuit}$ & ${\color{red}\diamondsuit}$ & ${\color{red}\heartsuit}$ & ${\color{red}\heartsuit}$ \\ \hline
    9 & ${\color{red}\heartsuit}$ & ${\color{red}\heartsuit}$ & ${\color{red}\heartsuit}$ &     ${\color{red}\heartsuit}$ & ${\color{red}\heartsuit}$ & ${\color{red}\heartsuit}$ & ${\color{red}\heartsuit}$  & ${\color{red}\heartsuit}$ & ${\color{red}\heartsuit}$ & ${\color{red}\heartsuit}$ 
& ${\color{red}\heartsuit}$ & ${\color{red}\heartsuit}$  & $\clubsuit$ & $\clubsuit$ &  &  &  & ${\color{red}\diamondsuit}$ & ${\color{red}\diamondsuit}$ \\ \hline
    10 & ${\color{red}\heartsuit}$ & ${\color{red}\heartsuit}$  & ${\color{red}\heartsuit}$ &   ${\color{red}\heartsuit}$ & ${\color{red}\heartsuit}$ & ${\color{red}\heartsuit}$ & ${\color{red}\heartsuit}$ & ${\color{red}\heartsuit}$ & ${\color{red}\heartsuit}$ & ${\color{red}\heartsuit}$ 
& ${\color{red}\heartsuit}$ & ${\color{red}\heartsuit}$&  &  & $\clubsuit$ & $\clubsuit$ & $\clubsuit$ & $\clubsuit$ & $\clubsuit$ \\ \hline
    11 & ${\color{red}\heartsuit}$ & ${\color{red}\heartsuit}$ &   &    &  & $\clubsuit$ & $\clubsuit$ &   $\clubsuit$ & ${\color{red}\heartsuit}$ &${\color{red}\heartsuit}$ 
& ${\color{red}\heartsuit}$ & ${\color{red}\heartsuit}$&  &  & ${\color{red}\diamondsuit}$ & ${\color{red}\diamondsuit}$ & ${\color{red}\diamondsuit}$ & ${\color{red}\diamondsuit}$ & ${\color{red}\diamondsuit}$\\ \hline
    12 & ${\color{red}\heartsuit}$ & ${\color{red}\heartsuit}$ &   &    &  & $\clubsuit$ & $\clubsuit$ &   $\clubsuit$ &${\color{red}\heartsuit}$ & ${\color{red}\heartsuit}$  
& ${\color{red}\heartsuit}$ & ${\color{red}\heartsuit}$&  &  & ${\color{red}\diamondsuit}$ & ${\color{red}\diamondsuit}$ & ${\color{red}\diamondsuit}$ & ${\color{red}\diamondsuit}$ & ${\color{red}\diamondsuit}$\\ \hline
    13 &  &  &   &       &  & ${\color{red}\heartsuit}$ & ${\color{red}\heartsuit}$ &  ${\color{red}\heartsuit}$ &  $\clubsuit$ &$\clubsuit$ 
& $\clubsuit$ & $\clubsuit$&  & ${\color{red}\diamondsuit}$ &  & $\spadesuit$ &  &  $\bigstar$ &  $\bigstar$ \\ \hline
    14 &  &  & &  &  &  &  &  &  &   
&  & &  & ${\color{red}\heartsuit}$ &  & $\clubsuit$ &  & ${\color{red}\diamondsuit}$ & ${\color{red}\diamondsuit}$\\ \hline
    15 &  & ${\color{red}\heartsuit}$ & ${\color{red}\heartsuit}$ &      ${\color{red}\heartsuit}$ & ${\color{red}\heartsuit}$ &  &  &  &   &   
&  & &  & $\clubsuit$ &  & ${\color{red}\diamondsuit}$ &  & ${\color{red}\heartsuit}$ & ${\color{red}\heartsuit}$ \\ \hline
    16 &  &  &  &  &  &  &  &  &  &  
&  & &  &  & ${\color{red}\heartsuit}$ & ${\color{red}\heartsuit}$ & ${\color{red}\heartsuit}$ &  &  \\ \hline \hline
    21 &  & ${\color{red}\heartsuit}$ &  ${\color{red}\heartsuit}$ &    ${\color{red}\heartsuit}$ & ${\color{red}\heartsuit}$ &  &  &  &   &   
& &  &  &  &  &  &  & $\clubsuit$ & ${\color{red}\heartsuit}$\\ \hline
    22  &  & ${\color{red}\heartsuit}$ &   &   &  &  & ${\color{red}\heartsuit}$ &  &   &  
& &  & ${\color{red}\heartsuit}$ &  &  &  &  &  & ${\color{red}\heartsuit}$\\ \hline
    23  & & $\clubsuit$ &      &       &  &  & $\clubsuit$ &  &  &  
&  & ${\color{red}\heartsuit}$ &  &  &  &  &  &  & ${\color{red}\heartsuit}$\\ \hline
    24  &  & ${\color{red}\heartsuit}$ &         &       &  &  & ${\color{red}\heartsuit}$ &  &  &  
&  & &  &  &  &  &  &  & \\ \hline
    25  &  & ${\color{red}\heartsuit}$ &         &       &  &  & ${\color{red}\heartsuit}$ &  &   &  
&  &&  &  &  &  &  &  & \\ \hline
    26  & & ${\color{red}\heartsuit}$ &        & &  &  & ${\color{red}\heartsuit}$ &  &   &  
&  & ${\color{red}\heartsuit}$&  &  &  &  &  &  & ${\color{red}\heartsuit}$ 
\\ \hline
    27-29  &  &  &  & &  &  &  &  &   &  
&  &  &  & ${\color{red}\heartsuit}$ &  &  &  &  & \\ \hline
    30  &  &  &  &  &  &  &  &  &  &  
&  &  &  & ${\color{red}\heartsuit}$ & ${\color{red}\heartsuit}$ & ${\color{red}\heartsuit}$ & ${\color{red}\heartsuit}$ &  & \\ \hline
 31  &  &  &  &  &  &  &  &  &  &   
&  & &  & ${\color{red}\heartsuit}$ & ${\color{red}\heartsuit}$ & ${\color{red}\heartsuit}$ & ${\color{red}\heartsuit}$ &  & \\ \hline
    32  &  &  &  &  &  &  &  &  &  &  &
 & &  & ${\color{red}\heartsuit}$ & $\clubsuit$ & $\clubsuit$ & $\clubsuit$ &  & \\ \hline
    33  &  &  &  &  &  &  &  &  &  & & 
 &  &  & ${\color{red}\heartsuit}$ & ${\color{red}\heartsuit}$ & ${\color{red}\heartsuit}$ & ${\color{red}\heartsuit}$ &  & \\ \hline
    34  &  &  &  &  &  &  &  &  &  &  &
 & &  & ${\color{red}\heartsuit}$ & ${\color{red}\heartsuit}$ & ${\color{red}\heartsuit}$ & ${\color{red}\heartsuit}$ &  & \\ \hline
    35-57  &  &  &  &  &  &  &  &  &  &  &
 & &  &  &  & ${\color{red}\heartsuit}$ &  &  & \\ \hline \hline
    62  &  &  &  &  &  &  &  &  &  &  &
  & &  &  &  & ${\color{red}\heartsuit}$ &  &  &  \\ \hline
    63-65  &  &  &  &   &  &  &  &  &  ${\color{red}\heartsuit}$ &  &
  & &  &  &  & $\clubsuit$ &  &  &  \\ \hline
    66 & & & & &  & ${\color{red}\heartsuit}$ & ${\color{red}\heartsuit}$ &   ${\color{red}\heartsuit}$ &  & &
 &  &  &  &  & $\clubsuit$ &  &  &  \\ \hline 
    67 &  &  &  ${\color{red}\heartsuit}$ & & ${\color{red}\heartsuit}$ & $\clubsuit$ & $\clubsuit$ &   ${\color{red}\diamondsuit}$ &  & &
 &  &  &  &  & $\spadesuit$ &  &  &   \\ \hline 
    68 & ${\color{red}\heartsuit}$ & $\clubsuit$ & ${\color{red}\heartsuit}$ & ${\color{red}\heartsuit}$ &              ${\color{red}\diamondsuit}$ & ${\color{red}\diamondsuit}$ & ${\color{red}\heartsuit}$ &  ${\color{red}\heartsuit}$  &$\clubsuit$ & ${\color{red}\heartsuit}$ &
 ${\color{red}\heartsuit}$ & ${\color{red}\heartsuit}$ & ${\color{red}\heartsuit}$ &  &  & ${\color{red}\heartsuit}$ &  & ${\color{red}\heartsuit}$ & ${\color{red}\heartsuit}$\\ \hline 
    69 & ${\color{red}\heartsuit}$ & $\clubsuit$ & ${\color{red}\diamondsuit}$ &         $\clubsuit$ & $\clubsuit$ & ${\color{red}\diamondsuit}$ & $\clubsuit$ & ${\color{red}\diamondsuit}$ &  ${\color{red}\diamondsuit}$ & ${\color{red}\diamondsuit}$ &
${\color{red}\heartsuit}$ & ${\color{red}\heartsuit}$ & $\spadesuit$ & $\spadesuit$ & $\spadesuit$ & $\spadesuit$ & ${\color{red}\heartsuit}$ & ${\color{red}\heartsuit}$ & ${\color{red}\heartsuit}$ \\ \hline 
    70 & ${\color{red}\heartsuit}$ & $\clubsuit$ & ${\color{red}\diamondsuit}$   & ${\color{red}\diamondsuit}$ &        $\clubsuit$ & $\clubsuit$ & ${\color{red}\diamondsuit}$ &  ${\color{red}\diamondsuit}$  & $\clubsuit$ &${\color{red}\diamondsuit}$ &
${\color{red}\heartsuit}$ & ${\color{red}\heartsuit}$ & ${\color{red}\diamondsuit}$ & ${\color{red}\diamondsuit}$ & ${\color{red}\heartsuit}$ & ${\color{red}\heartsuit}$ & ${\color{red}\heartsuit}$ & ${\color{red}\heartsuit}$ & ${\color{red}\heartsuit}$\\ \hline
    71 & ${\color{red}\heartsuit}$ & ${\color{red}\heartsuit}$ &         $\clubsuit$ &  ${\color{red}\heartsuit}$ & ${\color{red}\heartsuit}$ & $\clubsuit$ & ${\color{red}\heartsuit}$ & $\clubsuit$ & $\clubsuit$ & $\clubsuit$ &
${\color{red}\heartsuit}$ & ${\color{red}\heartsuit}$& $\clubsuit$ & $\clubsuit$ & ${\color{red}\heartsuit}$ & ${\color{red}\heartsuit}$ & ${\color{red}\heartsuit}$ & ${\color{red}\heartsuit}$ & ${\color{red}\heartsuit}$ \\ \hline
    72 & ${\color{red}\heartsuit}$ & ${\color{red}\heartsuit}$ & ${\color{red}\heartsuit}$ & ${\color{red}\heartsuit}$ & ${\color{red}\heartsuit}$ & ${\color{red}\heartsuit}$ & ${\color{red}\heartsuit}$ & ${\color{red}\heartsuit}$ & ${\color{red}\heartsuit}$ & ${\color{red}\heartsuit}$ &
${\color{red}\heartsuit}$ & ${\color{red}\heartsuit}$ & ${\color{red}\heartsuit}$ & ${\color{red}\heartsuit}$ & ${\color{red}\heartsuit}$ & ${\color{red}\heartsuit}$ & ${\color{red}\heartsuit}$ & ${\color{red}\heartsuit}$ & ${\color{red}\heartsuit}$\\ \hline 
\end{tabular}
\caption{\em Primary intermediates for SV40 and neighbouring regions of the phase space }
\label{SV40table1}
\end{scriptsize}
\end{table}
\end{landscape}
\section*{Glossary}

{\bf Assembly tree}: representation of all possible pathways in the building-up of a viral capsid through the association of a single capsomer at a time; usually allows for ramifications or branches which ultimately converge when the final capsid is formed. A branchless tree is called a linear assembly tree.

{\bf Capsid}: protein shell that encloses the nucleic acid of a virus particle. Commonly exhibits icosahedral symmetry and may be itself enclosed in an envelope (although not in the case of SV40).
The capsid is built up of protein subunits (integer multiples of 60) that self-assemble in a pattern typical of a particular virus. 

{\bf Capsomer}: regular polygonal grouping of proteins seen on the surface of viruses, consisting of usually 3 (trimer), 5 (pentamer), or 6 (hexamer) of the capsid proteins.

{\bf C-terminal arm}: One end of the polypeptide chain of a protein is called the C-terminus (the other end being the N-terminus). C-terminal arms are bonds obtained by binding of the C-terminal arm extension of one protein to the polypeptide chain of another. 

{\bf Icosahedral symmetry}: a body with icosahedral symmetry possesses a number of axes about which it may be rotated to give identical appearances. These are six 5-fold, ten 3-fold  and fifteen 2-fold axes of symmetry.

{\bf Papovaviridae}: family of DNA viruses including papilloma, polyoma and simian vacuolating
virus (SV40). These viruses are small, non-enveloped and mainly infect mammals. They may be linked to some forms of cancer.

{\bf (pseudo-)$T=7$ capsid}: capsid with a surface structure in which the locations of the capsomers follow geometries with T-number 7, but the capsomers do not exhibit the required symmetry properties, having fewer protein building blocks than predicted by the corresponding geometry with T-number 7.

{\bf Quasi-equivalence}: extent of similarity between structurally unique environments occupied by chemically identical protein subunits in viral capsids.

{\bf Tiling}: partition of a space into a countable number of building blocks called tiles that cover the space without gaps or overlaps. It requires matching rules indicating how tiles are fitted together. These matching rules may be realised through decoration of tiles. Unlike Penrose tiles, which tile the plane aperiodically, the tiles used here provide a tessellation of a closed surface with icosahedral symmetry.

{\bf T-number}: triangulation number, which corresponds to the number of unique quasi-equivalent environments present in a given icosahedral surface lattice. It is given by $T=h^2+hk+k^2$ where $h$ and $k$ are nonnegative integers and have no common factors, so that $T$ is in the sequence 1,3,4,7,... SV40 corresponds to $h=2,k=1$, hence $T=7$ dextro (d) [the case $h=1,k=2$ yields $T=7$ laevo (l)]. The triangulation number classifies the geometries of triangulations with icosahedral symmetry and the terminology triangulation number refers to the fact that it can be interpreted as the number of triangles in the triangulation of one of the twenty triangular faces of the icosahedron. 

{\bf Vertex star}: in tiling theory, a set of tiles sharing the same vertex. Distinct vertex stars form the vertex atlas of the tiling. 

{\bf Viral Tiling Theory}: A theory that encodes the surface structures of viral capsids in tessellations such that each tile in the tessellation represents a dimer- or a trimer-interaction between proteins located in the corners of the tiles. 
\end{document}